\documentclass[aip,pof,reprint]{revtex4-1}

\usepackage{amsmath}
\usepackage{math_pack}
\usepackage{graphicx}
\usepackage{color}
\usepackage{natbib}
\newcommand\omegab{\boldsymbol{\omega}}
\newcommand\taub{\boldsymbol{\tau}}

\providecommand\upi{\pi}

\begin{document}

%\begin{frontmatter}

\title{The effect of polydispersity in a turbulent channel flow laden with finite-size particles} %Title of paper

%%%%%% Group authors per affiliation:
%%%%\author{Elsevier\fnref{myfootnote}}
%%%%\address{Radarweg 29, Amsterdam}
%%%%\fntext[myfootnote]{Since 1880.}

%%%%\author[mymainaddress]{Walter Fornari\corref{mycorrespondingauthor}}
%%%%\cortext[mycorrespondingauthor]{Corresponding author}
%%%%\ead{fornari@mech.kth.se}
%%%%%\ead[url]{www.elsevier.com}
%%%%
%%%%\author[mysecondaryaddress]{Francesco Picano}
%%%%%\ead{support@elsevier.com}
%%%%
%%%%\author[mymainaddress]{Luca Brandt}
%%%%
%%%%\address[mymainaddress]{SeRC and Linn\'e FLOW Centre, KTH Mechanics, SE-100 44 Stockholm, Sweden}
%%%%\address[mysecondaryaddress]{Department of Industrial Engineering, University of Padova, Via Venezia 1, 35131, Padova, Italy}

\author{W.~Fornari}
\affiliation{SeRC and Linn\'e FLOW Centre, KTH Mechanics, SE-100 44 Stockholm, Sweden}

\author{F.~Picano}
\affiliation{Department of Industrial Engineering, University of Padova, Via Venezia 1, 35131, Padova, Italy}

\author{L.~Brandt}
\affiliation{SeRC and Linn\'e FLOW Centre, KTH Mechanics, SE-100 44 Stockholm, Sweden}

\begin{abstract}
We study turbulent channel flows of monodisperse and polydisperse suspensions of finite-size spheres by means of 
Direct Numerical Simulations using an immersed boundary method to account for the dispersed phase. 
Suspensions with 3 different Gaussian distributions of particle radii are considered (i.e.\ 3 different standard 
deviations). The distributions are centered on the reference particle radius of the monodisperse suspension. In the 
most extreme case, the radius of the largest particles is 4 times that of the smaller particles. We consider two 
different solid volume fractions, $2\%$ and $10\%$. We find that for all polydisperse cases, both fluid and particles 
statistics are not substantially altered with respect to those of the monodisperse case. Mean streamwise 
fluid and particle velocity profiles are almost perfectly overlapping. Slightly larger differences are found for 
particle velocity fluctuations. These increase close to the wall and decrease towards the centerline as the standard 
deviation of the distribution is increased. Hence, the behavior of the suspension is mostly governed by excluded 
volume effects regardless of particle size distribution (at least for the radii here studied). Due to turbulent 
mixing, particles are uniformly distributed across the channel. However, smaller particles can penetrate more into 
the viscous and buffer layer and velocity fluctuations are therein altered. Non trivial results are presented for 
particle-pair statistics.
\end{abstract}

%%\begin{keyword}
%%Suspensions, particle-laden flows, particle/fluid flow
%%%\texttt{elsarticle.cls}\sep \LaTeX\sep Elsevier \sep template
%%%\MSC[2010] 00-01\sep  99-00
%%\end{keyword}

%\end{frontmatter}

%\linenumbers

\maketitle %\maketitle must follow title, authors, abstract and \pacs

\section{Introduction}

Particle laden flows are relevant in several industrial applications and many natural and environmental processes. Among these we 
recall the sediment transport in rivers, avalanches and pyroclastic flows, plankton in seas, planetesimals in accretion disks, as 
well as many oil industry and pharmaceutical processes. In most cases the carrier phase is a turbulent flow due to the high flow 
rates. 
However, due to the interaction between particles and vortical structures of different sizes the turbulence properties can be 
substantially altered and the flow may even be relaminarized. Additionally, particles may differ in density, shape, size and 
stiffness. The prediction of the suspension rheological behavior is hence a complex task. 

Interesting and peculiar rheological properties can be observed already in the viscous and low-speed laminar regimes, and for 
suspensions of monodispersed rigid spheres. Depending for example on the shear rate and on particle concentration, 
suspensions can exhibit shear thinning or thickening, jamming (at high volume fractions), and the generation of high effective 
viscosities and normal stress differences \cite{stickel2005,morris2009,wagner2009}. More generally, due to the dispersed solid 
phase, the fluid response to the local deformation rate is altered and the resulting suspension effective viscosity $\mu_e$ differs 
from that of the pure fluid $\mu$\cite{guazz2011,einstein1906,einstein1911,batchelor1970}. 
In laminar flows, when the particle Reynolds number $Re_a$ becomes non negligible, the symmetry of the particle pair trajectories 
is broken and the microstructure becomes anisotropic. This leads to macroscopical behaviors such as shear-thickening and the occurrence 
of normal stress differences\cite{Morrispof08,picano2013,Morris2014}. Recently, it was also shown that in simple shear flows, the effective 
viscosity $\mu_e$ depends non-monotonically on the system confinement (i.e. the gap size in a Couette flow). In particular, minima of 
$\mu_e$ are observed when the gap size is approximately an integer number of particle diameters, due to the formation of stable particle 
layers with low momentum exchange across layers\cite{fornariPRL}. 
Finally, in the Bagnoldian or highly inertial regime the effective viscosity $\mu_e$ increases linearly with shear rate due to augmented 
particle collisions \cite{bagnold1954}.

When particles are dispersed in turbulent flows, the dynamics of the fluid phase can be substantially modified. Already in the 
transition from the laminar to the turbulent regime, the presence of the solid phase may either increase or reduce the 
critical Reynolds number above which the transition occurs. Different groups\cite{matas2003,yu2013} studied for example, the 
transition in a turbulent pipe flow laden with a dense suspension of particles. They found that transition depends upon the pipe to particle 
diameter ratios and the volume fraction. For smaller neutrally-buoyant particles they observed that the critical Reynolds number increases 
monotonically with the solid volume fraction $\phi$ due to the raise in effective viscosity. On the other hand, for larger particles it was 
found that transition shows a non-monotonic behavior which cannot be solely explained in terms of an increase of the effective viscosity 
$\mu_e$. Concerning transition in dilute suspensions of finite-size particles in plane channels, it was shown that the critical Reynolds number above 
which turbulence is sustained, is reduced\cite{lashg2015,loisel2013}. At fixed Reynolds number and solid volume fraction, also the initial 
arrangement of particles was observed to be important to trigger the transition.\\ 
For channel flows laden with solid spheres, three different regimes have been identified for a wide range of solid volume fractions $\phi$ 
and bulk Reynolds numbers $Re_b$\cite{lashgari2014}. These are laminar,turbulent and inertial shear-thickening regimes and in each case, the 
flow is dominated by different components of the total stress: viscous, turbulent or particle stresses.

In the fully turbulent regime, most of the previous studies have focused on dilute or very dilute suspensions of particles smaller than the 
hydrodynamic scales and heavier than the fluid. In the one-way coupling regime \cite{balach-rev2010} (i.e.\ when the solid phase has a negligible 
effect on the fluid phase), it has been shown that particles migrate from regions of high to low turbulence intensities \cite{reeks1983}. This 
phenomenon is known as turbophoresis and it is stronger when the turbulent near-wall characteristic time and the particle inertial time scale are 
similar \cite{soldati2009}. In these inhomogeneous flows, Sardina et al.\cite{sardina2011,sardina2012} also observed small-scale clustering that 
together with turbophoresis leads to the formation of streaky particle patterns \cite{sardina2011}. When the solid mass fraction is high and 
back-influences the fluid phase (i.e. in the two-way coupling regime), turbulence modulation has been observed\cite{kulick1994,zhao2010}. The 
turbulent near-wall fluctuations are reduced, their anisotropy increases and eventually the total drag is decreased.

In the four-way coupling regime (i.e.\ dense suspensions for which particle-particle interactions must be considered), it was shown that 
finite-size particles slightly larger than the dissipative length scale increase the turbulent intensities and the Reynolds stresses \cite{pan1996}. 
Particles are also found to preferentially accumulate in the near-wall low-speed streaks. This was also observed in open channel 
flows laden with heavy finite-size particles \cite{kida2013}.\\
On the contrary, for turbulent channel flows of denser suspensions of larger particles (with radius of about $10$ plus units), it was 
found that the large-scale streamwise vortices are attenuated and that the fluid streamwise velocity fluctuation is reduced\cite{shao2012,picano2015}. 
The overall drag increases as the volume fraction is increased from $\phi=0\%$ up to $20\%$. As $\phi$ is increased, turbulence is progressively reduced 
(i.e. lower velocity fluctuation intensities and Reynolds shear stresses). However, particle induced stresses show the opposite behavior with $\phi$, 
and at the higher volume fraction they are the main responsible for the overall increase in drag\cite{picano2015}. Recently, Costa et 
al.\cite{costa2016} showed that if particles are larger than the smallest turbulent scales, the suspension deviates from the continuum 
limit. The effective viscosity alone is not sufficient to properly describe the suspension dynamics which is instead altered by the 
generation of a near-wall particle layer with significant slip velocity. 

As noted by Prosperetti \cite{prosp2015}, however, results obtained for solid to fluid density ratios $R=\rho_p/\rho_f=1$ and for spherical particles, 
cannot be easily extrapolated to other cases (e.g. when $R > 1$). This motivated researchers to investigate turbulent channel flows with different types 
of particles. For example, in an idealized scenario where gravity is neglected, we studied the effects of varying independently 
the density ratio $R$ at constant $\phi$, or both $R$ and $\phi$ at constant mass fraction, on both the rheology and the turbulence\cite{fornariPOF}. 
%the solid mass fraction or the density ratio $R$ on both the rheology and the turbulence\cite{fornariPOF}. 
We found that the influence of the 
density ratio $R$ on the statistics of both phases is less important than that of an increasing volume fraction $\phi$. However, for moderately high 
values of the density ratio ($R \sim 10$) we observed an inertial shear-induced migration of particles towards the core of the channel. Ardekani et 
al.\cite{ardekani2016} studied instead a turbulent channel flow laden with finite-size neutrally buoyant oblates. They showed that due to the peculiar particle 
shape and orientation close to the channel walls, there is clear drag reduction with respect to the unladen case.

In the present study we consider again finite-size neutrally buoyant spheres and explore the effects of polydispersity. Typically, it is 
very difficult in experiments to have suspension of precisely monodispersed spheres (i.e. with exactly the same diameter). On the other hand, direct numerical simulations 
(DNS) of particle laden flows are often limited to monodisperse suspensions. Hence, we decide to study turbulent channel flows 
laden with spheres of different diameters. Trying to mimic experiments, we consider suspensions with Gaussian distributions of diameters. We study 3 different 
distributions with $\sigma_a/(2a)=0.02, 0.06$ and $0.1$, being $\sigma_a$ the standard deviation. For each case we have a total of 7 different species and the 
solid volume fraction $\phi$ is kept constant at $10\%$ (for each case the total number of particles is different). We then consider a more dilute case with 
$\phi=2\%$ and $\sigma_a/(2a)=0.1$. The reference spheres have radius of size $a=h/18$ where $h$ is the half-channel height. The statistics for 
all $\sigma_a/(2a)$ are compared to those obtained for monodisperse suspensions with same $\phi$. For 
all $\phi$, we find that even for the larger $\sigma_a/(2a)=0.1$ the results do not differ substantially from those of the monodisperse case. Slightly larger 
variations are found for particle mean and fluctuating velocity profiles. Therefore, rheological properties and turbulence modulation depend strongly on the 
overall solid volume fraction $\phi$ and less on the particle size distribution. We then look at probability density functions of particle velocities and 
mean-squared dispersions. For each species the curves are similar and almost overlapped. However, we identify a trend depending on the 
particle diameter. Finally, we study particle-pair statistics. We find that collision kernels between particles of different sizes (but equal concentration), 
resemble more closely those obtained for equal particles of the smaller size.

\section{Methodology}
\subsection{Numerical method}

In the present study we perform direct numerical simulations and use an immersed boundary method to account for the presence 
of the dispersed solid phase\cite{breugem2012,kempe2012}. The Eulerian fluid phase is evolved according to the incompressible
Navier-Stokes equations,
\begin{equation}
\label{div_f}
\div \vec u_f = 0
\end{equation}
\begin{equation}
\label{NS_f}
\pd{\vec u_f}{t} + \vec u_f \cdot \grad \vec u_f = -\frac{1}{\rho_f}\grad p + \nu \grad^2 \vec u_f + \vec f
\end{equation}
where $\vec u_f$, $\rho_f$, $p$ and $\nu=\mu/\rho_f$ are the fluid velocity, density, pressure and kinematic viscosity respectively ($\mu$ is
the dynamic viscosity). The immersed boundary force $\vec f$, models the boundary conditions at the moving particle surface. 
The particles centroid linear and angular velocities, $\vec u_p$ and $\vec \omegab_p$ are instead governed by the Newton-Euler Lagrangian 
equations,
\begin{align}
\label{lin-vel}
\rho_p V_p \td{\vec u_p}{t} &= \rho_f \oint_{\partial \mathcal{V}_p}^{} \vec \taub \cdot \vec n\, dS\\
\label{ang-vel}
I_p \td{\vec \omegab_p}{t} &= \rho_f \oint_{\partial \mathcal{V}_p}^{} \vec r \times \vec \taub \cdot \vec n\, dS
\end{align}
where $V_p = 4\upi a^3/3$ and $I_p=2 \rho_p V_p a^2/5$ are the particle volume and moment of inertia; $\vec \taub = -p \vec I + 2\mu \vec E$ 
is the fluid stress, with $\vec E = \left(\grad \vec u_f + \grad \vec u_f^T \right)/2$ the deformation tensor; $\vec r$ is the distance vector 
from the center of the sphere while $\bf{n}$ is the unity vector normal to the particle surface $\partial \mathcal{V}_p$. Dirichlet boundary 
conditions for the fluid phase are enforced on the particle surfaces as $\vec u_f|_{\partial \mathcal{V}_p} = \vec u_p + \vec \omegab_p \times \vec r$.\\
The fluid phase is evolved in the whole computational domain using a second order finite difference scheme on a staggered mesh. The time integration 
of both Navier-Stokes and Newton-Euler equations is performed by a third order Runge-Kutta scheme. A pressure-correction method is applied at each 
sub-step. Each particle surface is described by $N_L$ uniformly distributed Lagrangian points. 
The force exchanged by fluid and the particles is imposed on each $l-th$ Lagrangian point and is related to the Eulerian force field 
$\vec f$ by the expression $\vec f(\vec x) = \sum_{l=1}^{N_L} \vec F_l \delta_d(\vec x - \vec X_l) \Delta V_l$. In the latter $\Delta V_l$ 
represents the volume of the cell containing the $l-th$ Lagrangian point while $\delta_d$ is the Dirac delta. This force field is calculated 
through an iterative algorithm that ensures a second order global accuracy in space.\\ 
Particle-particle interactions are also considered. When the gap distance between two particles is smaller than twice the mesh size, lubrication models 
based on Brenner's and Jeffrey's asymptotic solutions \citep{brenner1961,jeffrey1982} are used to correctly reproduce the interaction between the particles of 
different sizes. A soft-sphere collision 
model is used to account for collisions between particles and between particles and walls. An almost elastic rebound is ensured with a restitution coefficient 
set at $0.97$. These lubrication and collision forces are added to the Newton-Euler equations. For more details and validations of the numerical code, the 
reader is referred to previous publications \cite{breugem2012,lambert2013,fornari2015}.

\subsection{Flow configuration}

\begin{figure}
\centering
\includegraphics[width=.60\textwidth]{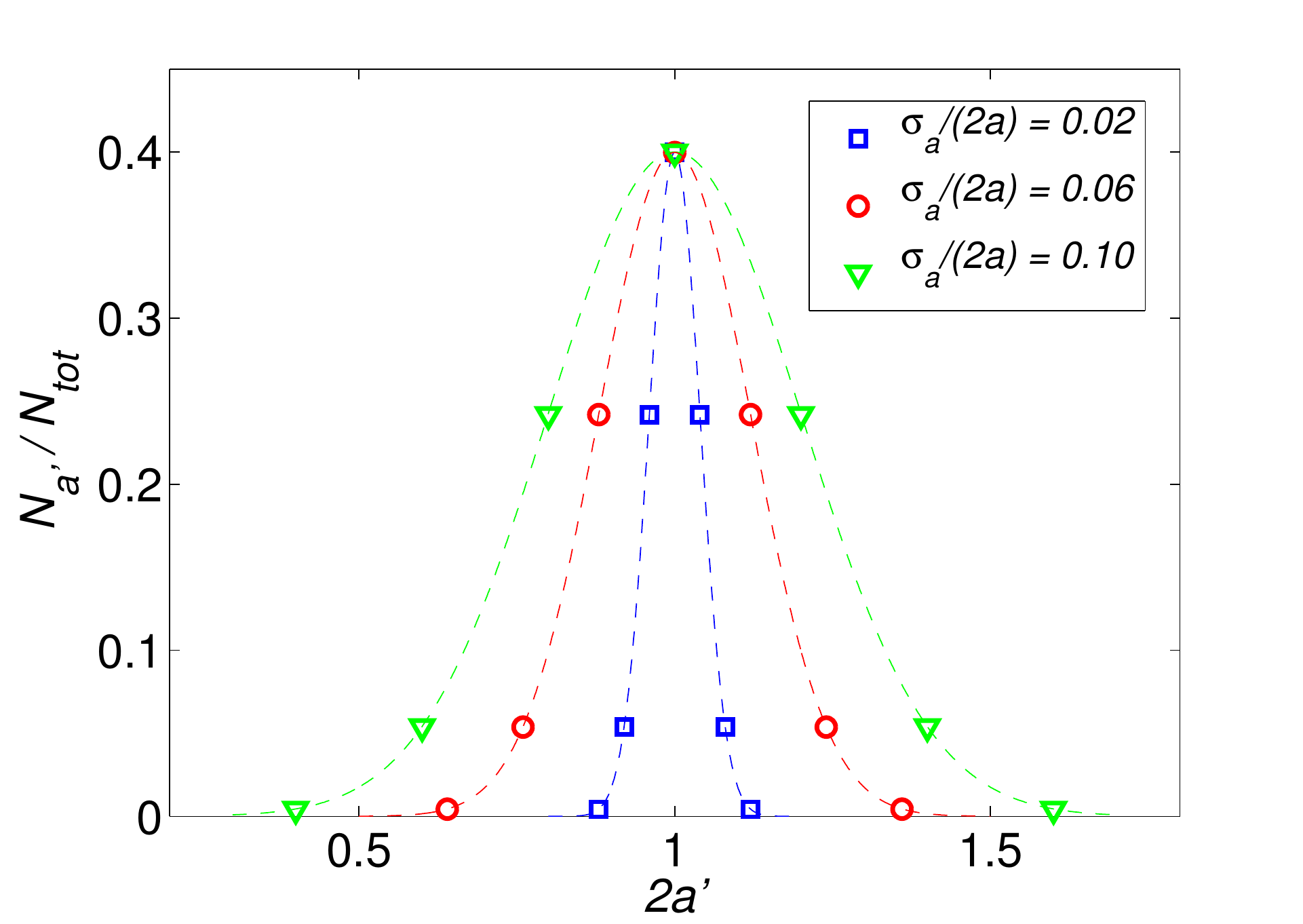}
\caption{Fraction of particles with radius $a'$, $N_{a'}/N_{tot}$, for each Gaussian distribution. \label{fig:imm1}}
\label{fig:snap}
\end{figure}

We consider a turbulent channel flow between two infinite flat walls located at $y=0$ and $y=2h$, where $y$ is the wall-normal direction while 
$x$ and $z$ are the streamwise and spanwise directions. The domain has size $L_x=6h$, $L_y=2h$ and $L_z=3h$ with periodic 
boundary conditions imposed in the streamwise and spanwise directions. A mean pressure gradient is imposed in the streamwise direction to ensure a fixed 
value of the bulk velocity $U_0$. The imposed bulk Reynolds number is equal to $Re_b=U_02h/\nu=5600$ and corresponds to a 
Reynolds number based on the friction velocity $Re_{\tau}=U_*h/\nu=180$ for the unladen case. The friction velocity is defined as 
$U_*=\sqrt{\tau_w/\rho_f}$, where $\tau_w$ is the stress at the wall. A staggered mesh of $1296\times432\times649$ grid points is used 
to discretize the domain. All results will be reported either in non-dimensional outer units (scaled by $U_0$ and $h$) or in inner units (with the superscript '+', 
using $U_*$ and $\delta_*=\nu/U_*$).

\begin{figure}
\centering
\includegraphics[width=.70\textwidth]{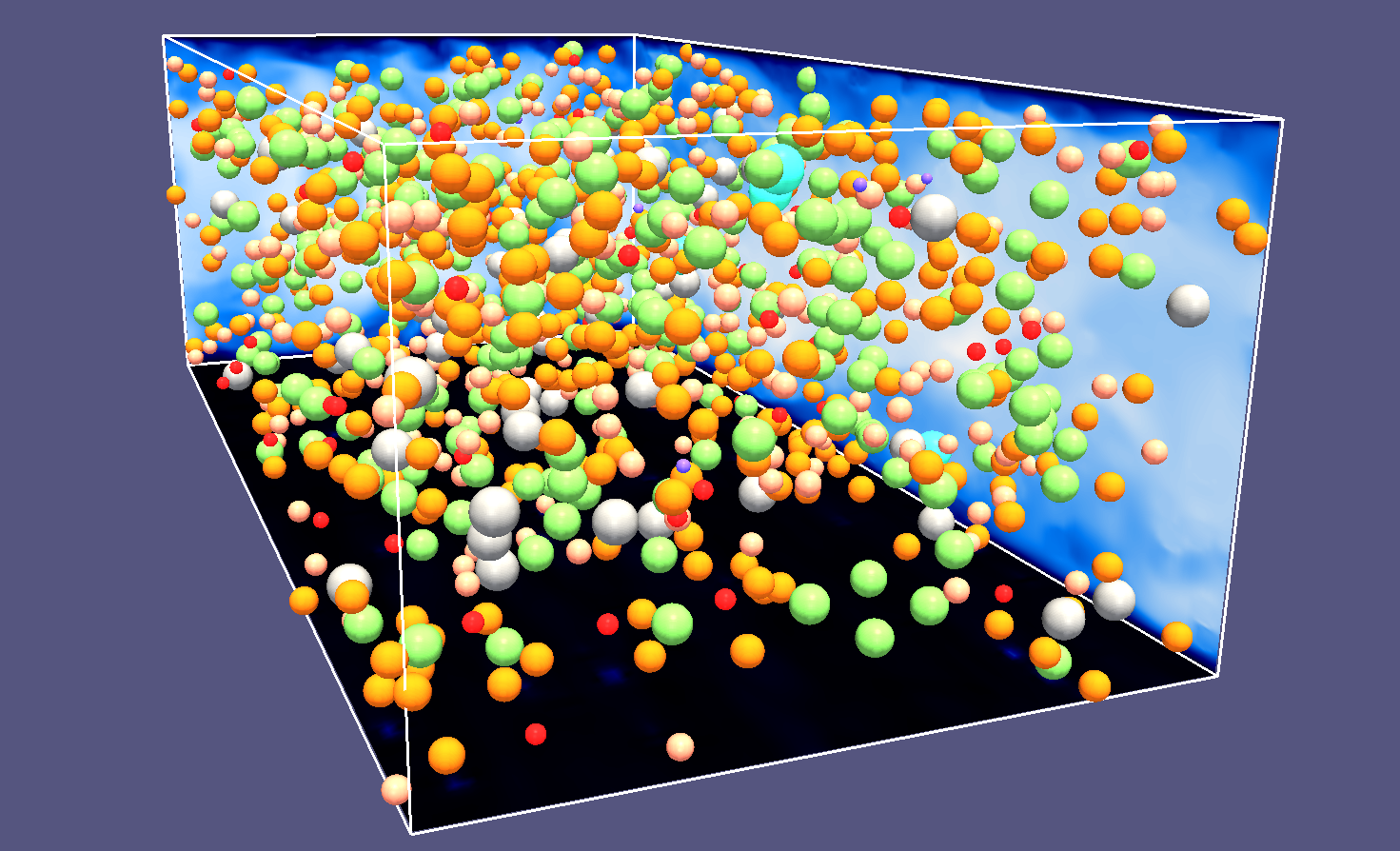}
\caption{Instantaneous snapshot of the instantaneous streamwise velocity on three orthogonal planes together with a fraction of particles for the case $\sigma_a/(2a)=0.1$. \label{fig:imm2}}
\label{fig:snap}
\end{figure}

The solid phase consists of non-Brownian, neutrally buoyant rigid spheres of different sizes. In particular we consider Gaussian distributions of particle radii with 
standard deviations of $\sigma_a/(2a)=0.02, 0.06$ and $0.1$. In figure~\ref{fig:imm1} we show for each $\sigma_a/(2a)$, the fraction of particles with radius $a'$, 
$N_{a'}/N_{tot}$ (with $N_{tot}$ the total number of spheres). For all cases, the reference spheres have a radius to channel half-width ratio fixed to $a/h=1/18$. 
The reference particles are discretized with $N_l=1721$ Lagrangian control points while their radii are $12$ Eulerian grid points long. 
In figure~\ref{fig:imm2} we display the instantaneous streamwise velocity on three orthogonal planes together with a fraction of the particles dispersed in the domain 
for $\sigma_a/(2a)=0.1$. In this extreme case, the size of the smallest and largest particles is $a'/a=0.4$ and $1.6$. These particles are hence substantially smaller/larger 
than our reference spheres.

The simulations start from the laminar Poiseuille flow for the fluid phase since we observe that the transition naturally occurs at the present moderately high Reynolds 
number due to the noise added by the particles. Particles are initially positioned randomly with velocity equal to the local fluid velocity. Statistics are collected 
after the initial transient phase. At first, we will compare results obtained for denser suspensions with solid volume fraction $\phi=10\%$ and different $\sigma_a/(2a)$, 
with those of the monodisperse case ($\sigma_a/(2a)=0$). We will then discuss the statistics obtained for $\phi=2\%$ and $\sigma_a/(2a)=0$ and $0.1$. The full set of simulations is 
summarized in table~\ref{tab:sim}.

\begin{table}
  \begin{center}
\def~{\hphantom{0}}
  \begin{tabular}{ccc}
     $\phi ($\%$)$  & $\sigma_a/(2a)$ & $N_p$    \\[3pt]
     $10$           &   $0$         & $5012$    \\
     $10$           &   $0.02$      & $4985$    \\
     $10$           &   $0.06$      & $4802$    \\
     $10$           &   $0.10$      & $4474$    \\
     $2$            &   $0$         & $1002$    \\
     $2$            &   $0.10$       & $896$    \\
  \end{tabular}
  \caption{Summary of the simulations performed ($N_p$ is the total number of particles).}
  \label{tab:sim}
  \end{center}
\end{table}

\section{Results}

\subsection{Fluid and particle statistics}

\begin{figure}
\centering
\includegraphics[width=.50\textwidth]{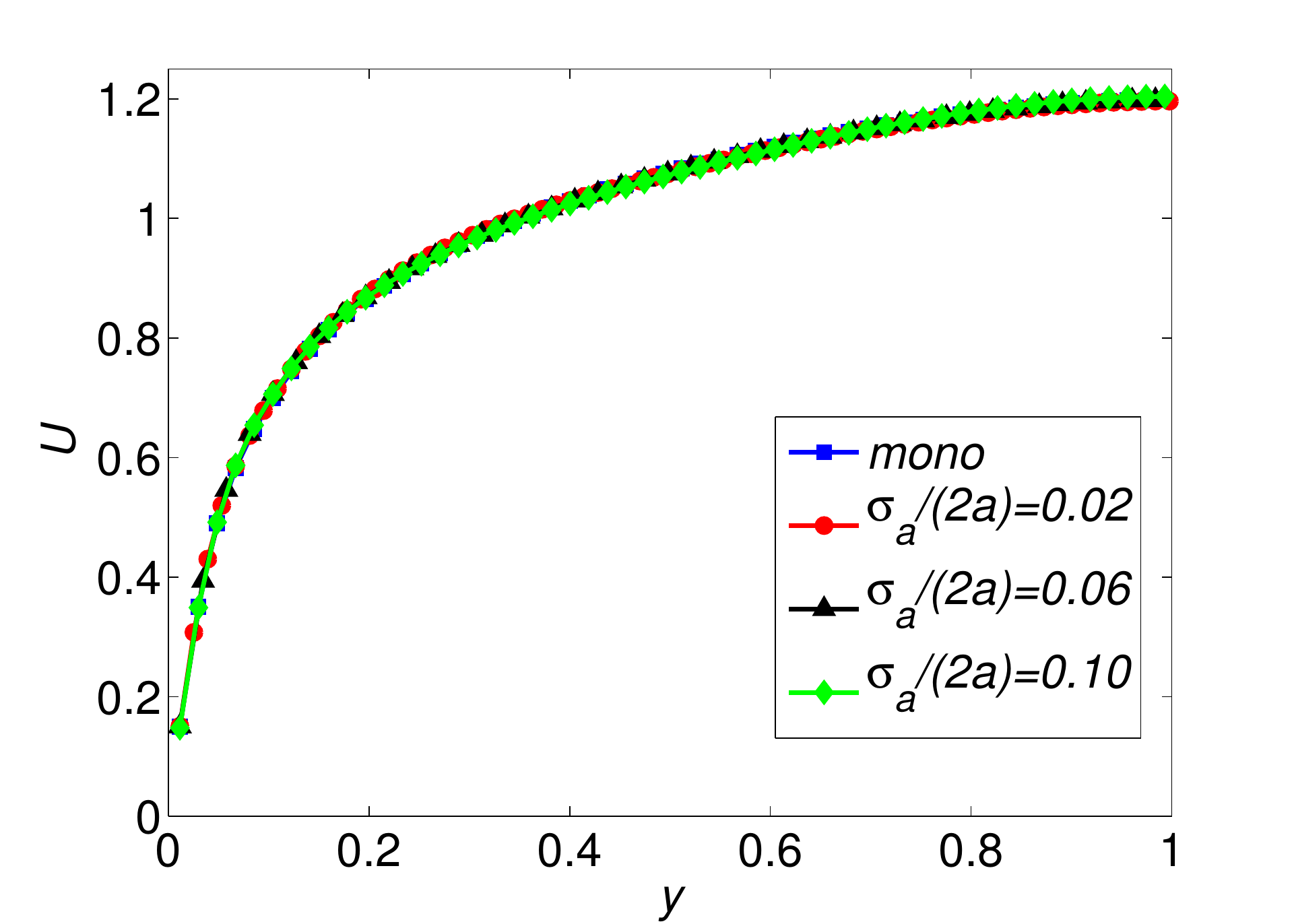}~~~~
\put(-185,110){{\large a)}}
{\includegraphics[width=.50\textwidth]{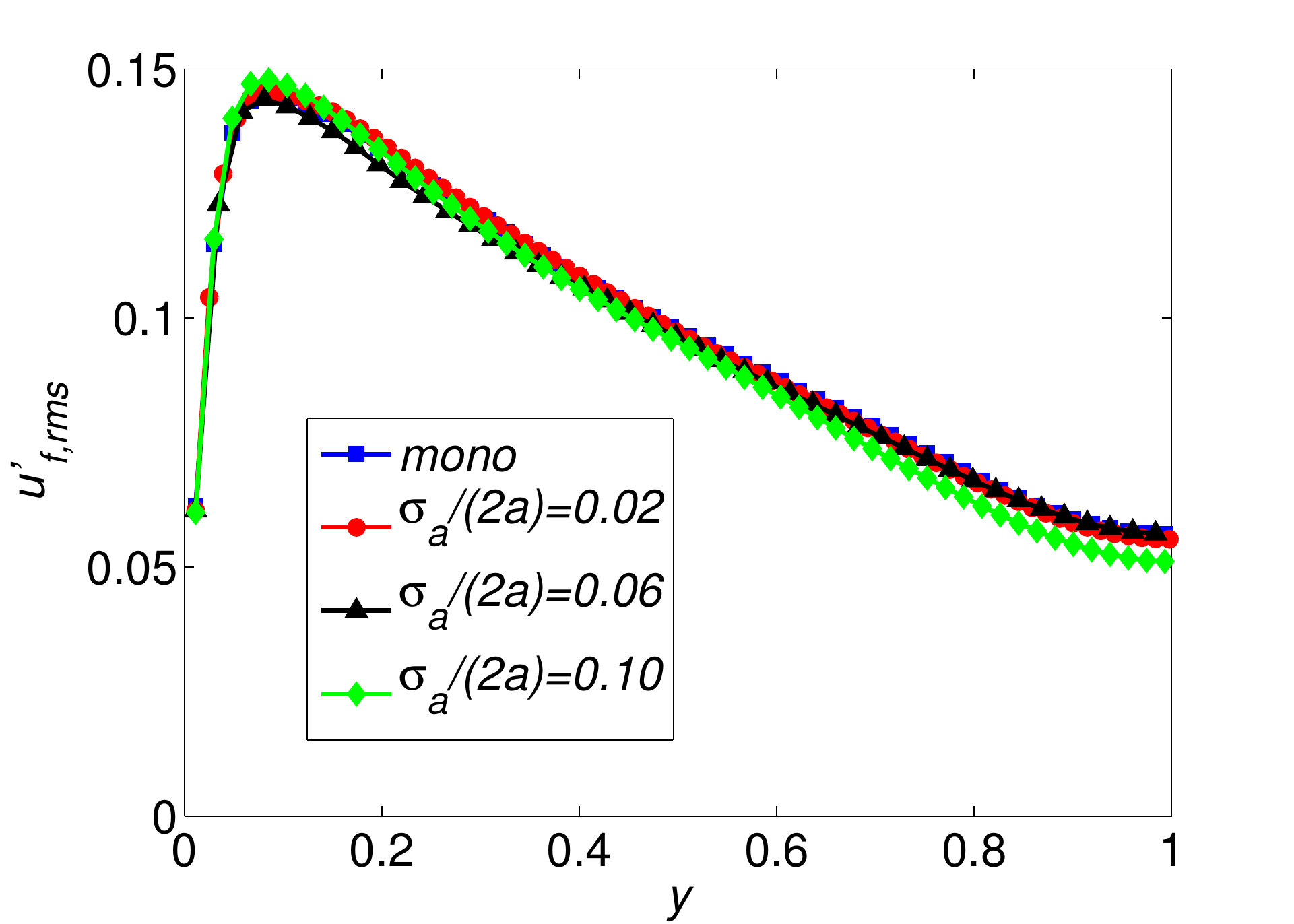}
\put(-180,110){{\large b)}}}
\includegraphics[width=.50\textwidth]{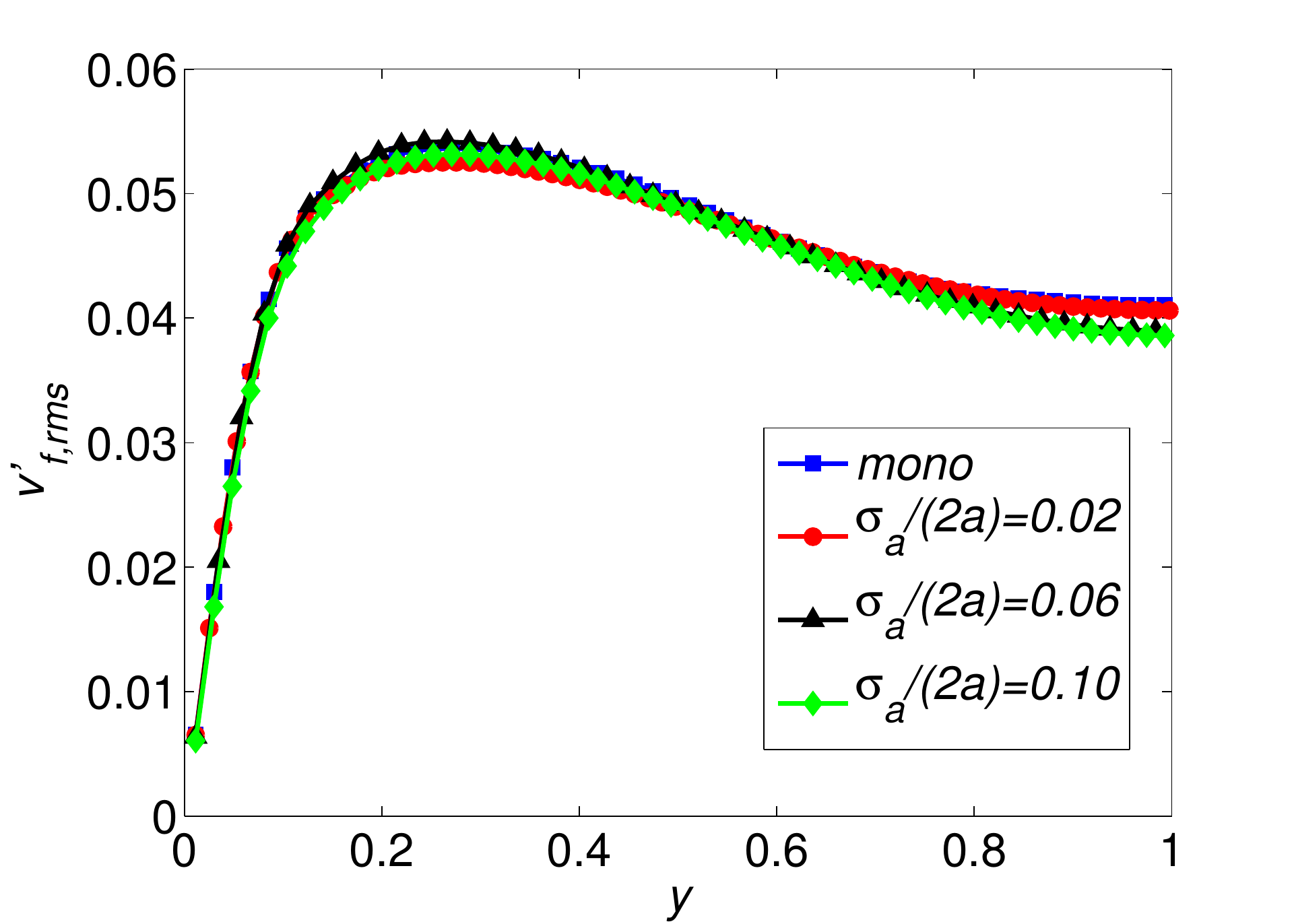}~~~~
\put(-185,110){{\large c)}}
{\includegraphics[width=.50\textwidth]{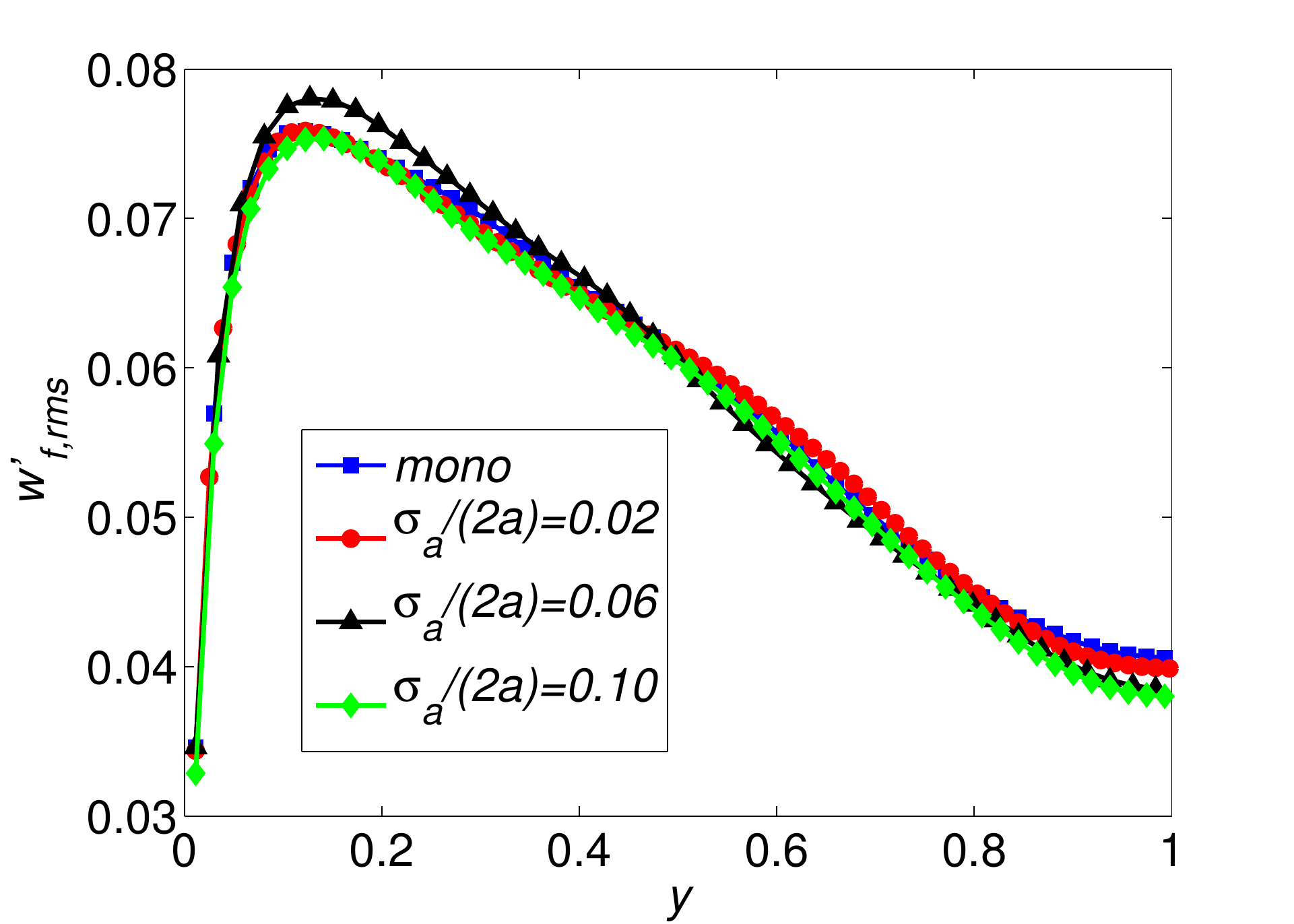}
\put(-180,110){{\large d)}}}
\caption{Mean fluid streamwise velocity profile (a) and fluid velocity fluctuations in the streamwise (b), wall-normal (c) and spanwise (d) 
directions for different standard deviations $\sigma_a/(2a)=0$, $0.02$, $0.06$, $0.1$.}
\label{fig:Uf}
\end{figure}

We show in figure~\ref{fig:Uf}$(a)$ the mean fluid streamwise velocity profiles in outer units, $U(y)$, for $\sigma_a/(2a)=0, 0.02, 0.06$ and $0.1$. We 
find that the profiles obtained for monodisperse and polydisperse suspensions overlap almost perfectly. No differences are observed 
even for the case with larger variance $\sigma_a/(2a)=0.1$. In figures~\ref{fig:Uf}$(b),(c),(d)$ we then show the profiles of streamwise, 
wall-normal and spanwise fluctuating fluid velocities, $u_{f,rms}', v_{f,rms}', w_{f,rms}'$. These profiles exhibit small variations and 
no precise trend (as function of $\sigma_a/(2a)$) can be identified. The larger variations between the cases are found close to the wall, 
$y \in (0.1;0.2)$, where the maximum intensity of the velocity fluctuations is found, and at the centerline. In the latter location, we notice 
that fluctuations are always smaller for $\sigma_a/(2a)=0.1$. 
In this case, many particles are substantially larger than the reference ones with $a'/a=1$. Around the centerline these move almost 
undisturbed therefore inducing slightly smaller fluid velocity fluctuations.

\begin{figure}
\centering
\includegraphics[width=.50\textwidth]{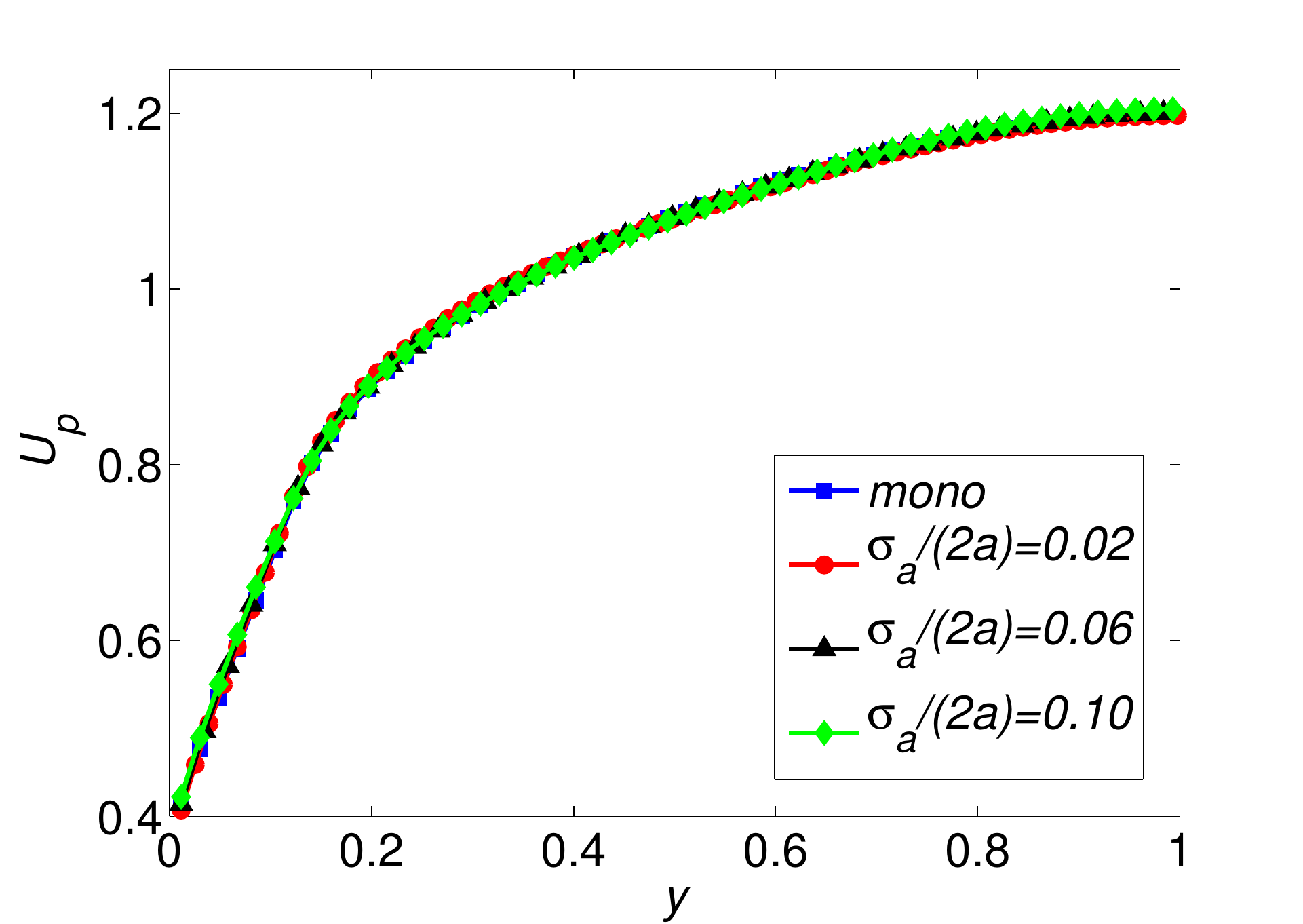}
\put(-185,110){{\large a)}}\\
\includegraphics[width=.50\textwidth]{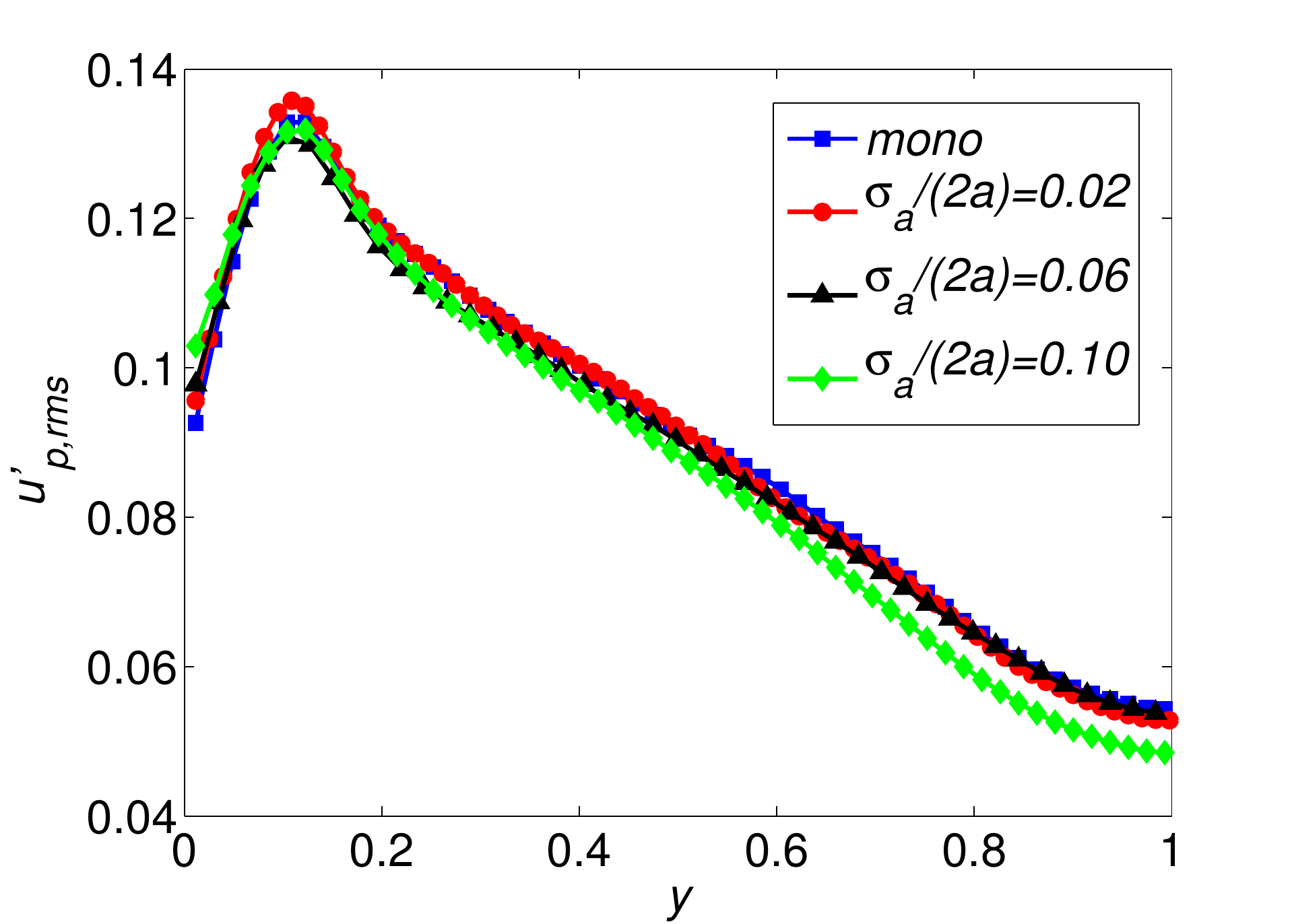}~~~~
\put(-185,110){{\large b)}}
{\includegraphics[width=.50\textwidth]{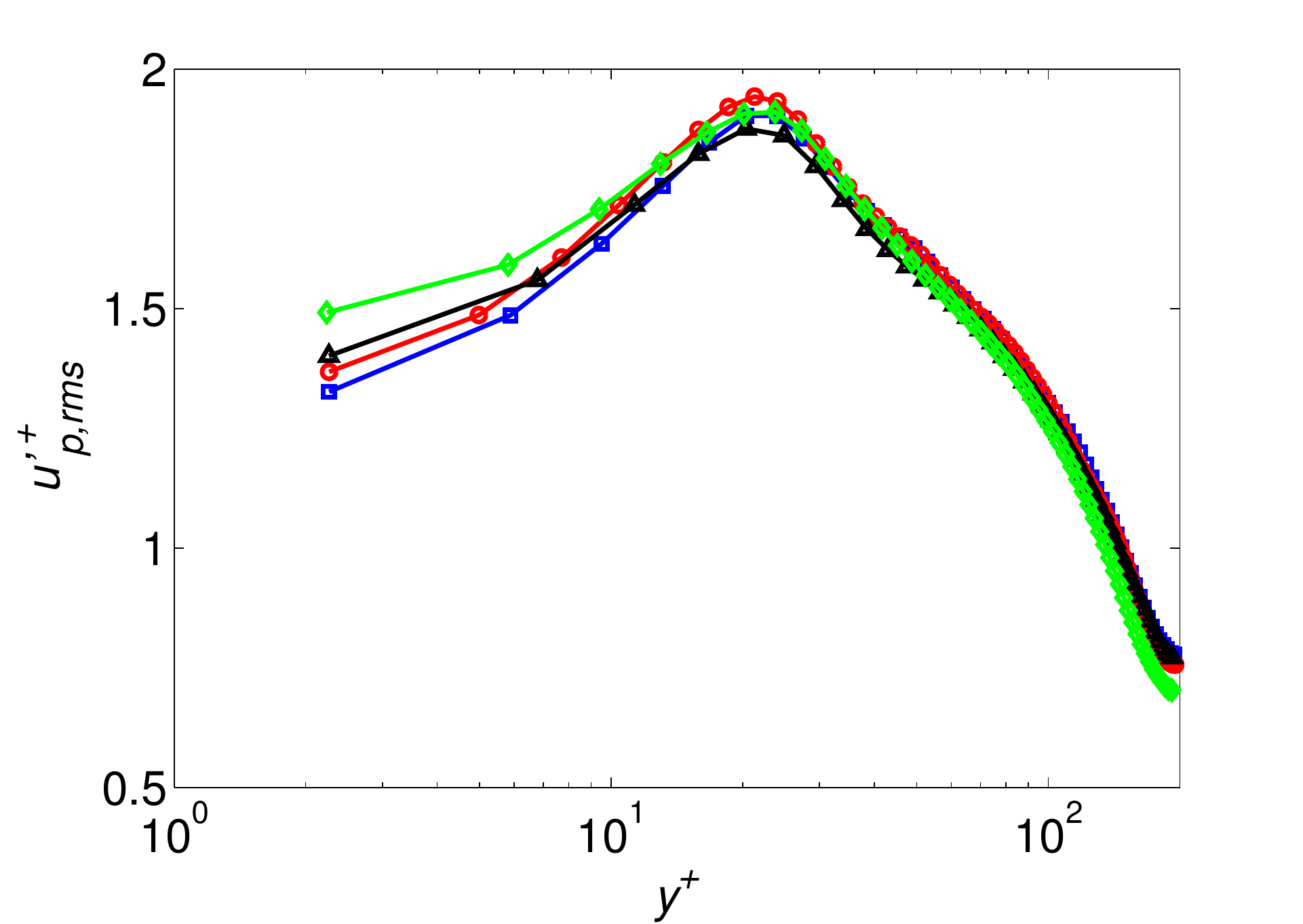}
\put(-180,110){{\large c)}}}
\includegraphics[width=.50\textwidth]{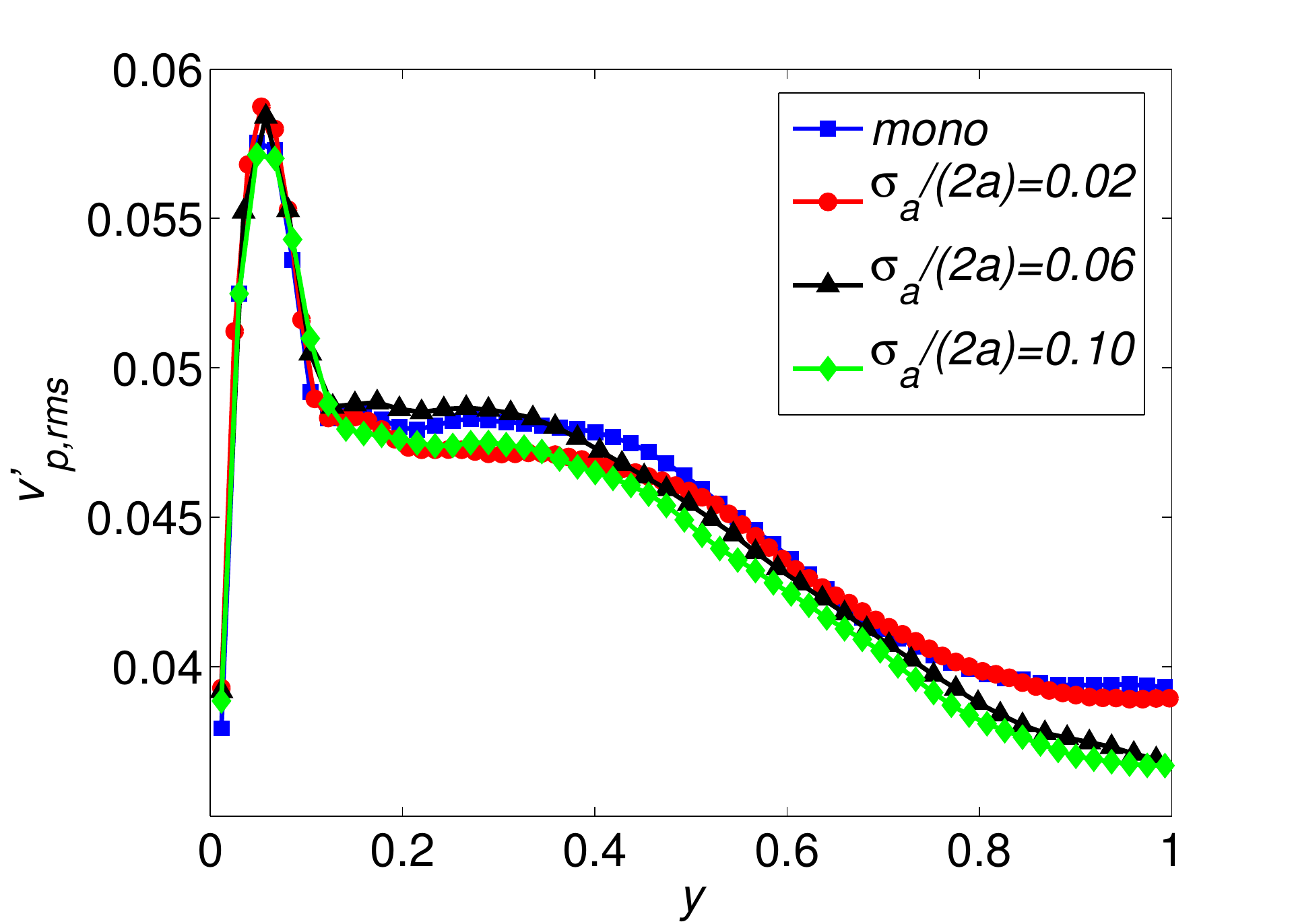}~~~~
\put(-185,110){{\large d)}}
{\includegraphics[width=.50\textwidth]{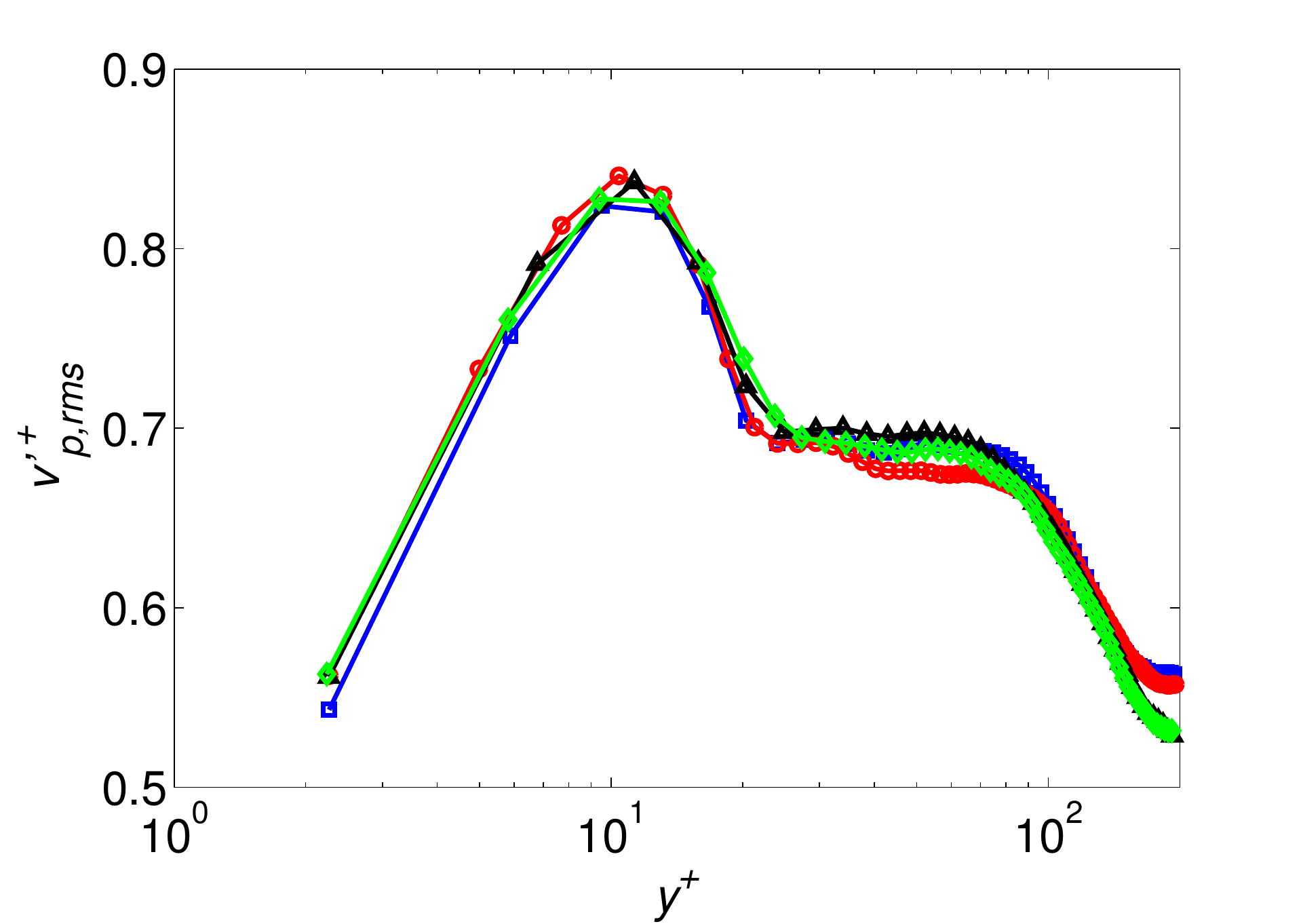}
\put(-180,110){{\large e)}}}
\includegraphics[width=.50\textwidth]{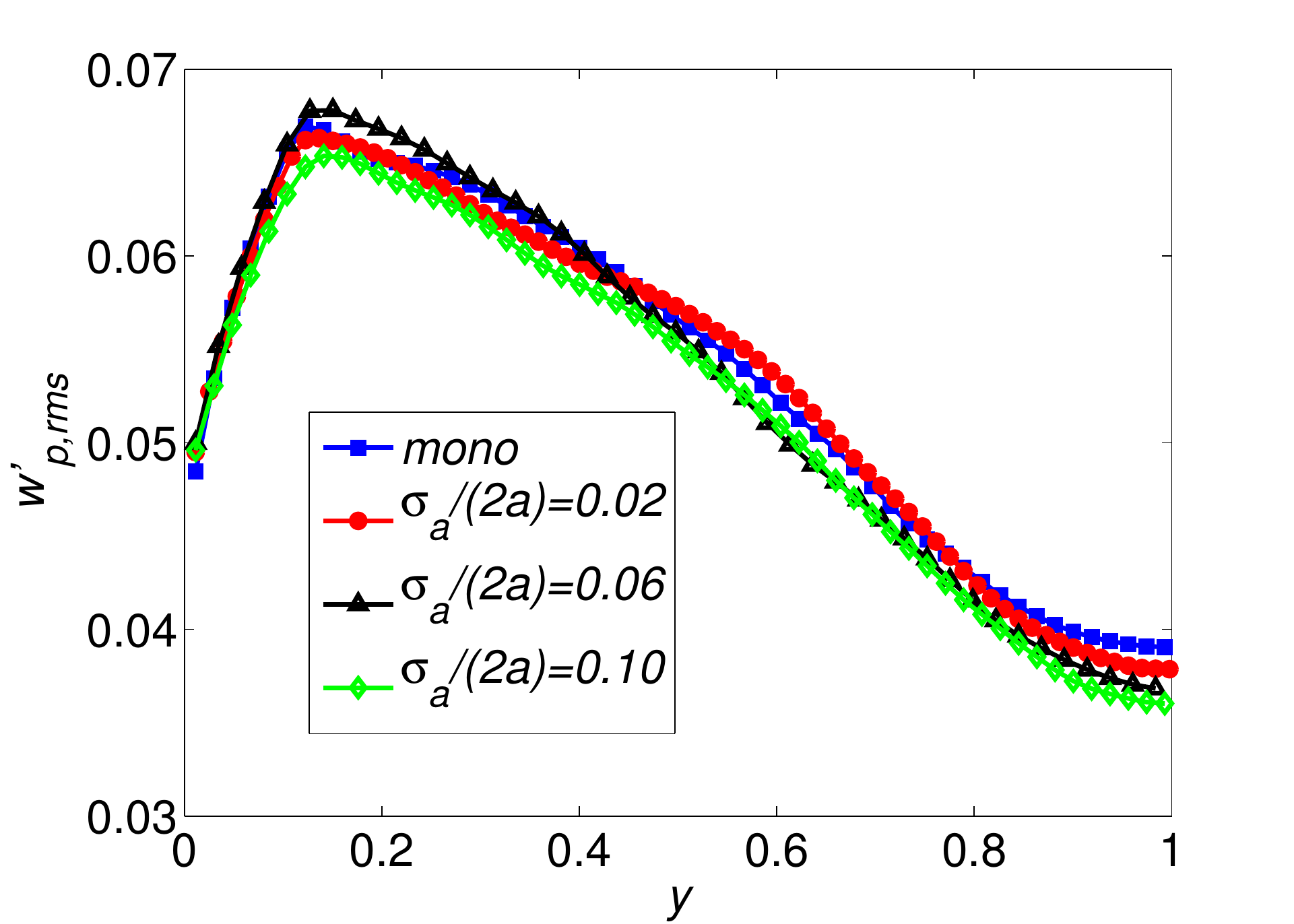}~~~~
\put(-185,110){{\large f)}}
{\includegraphics[width=.50\textwidth]{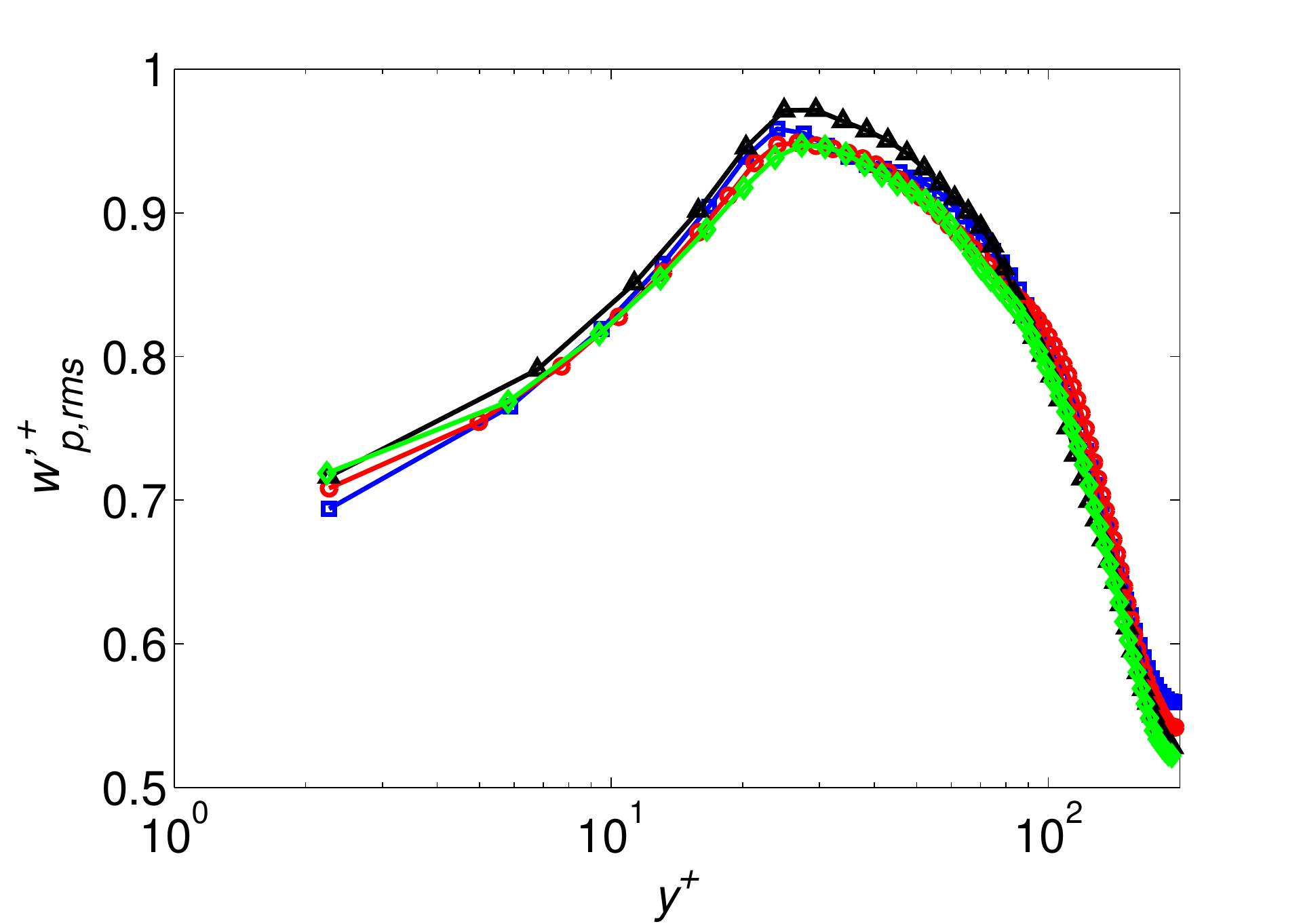}
\put(-180,110){{\large g)}}}
\caption{Mean particle streamwise velocity profile (a) and particle velocity fluctuations in outer and inner units, in the streamwise (b)-(c), 
wall-normal (d)-(e) and spanwise (f)-(g) directions, for different standard deviations $\sigma_a/(2a)=0$, $0.02$, $0.06$, $0.1$.}
\label{fig:Up}
\end{figure}

The mean particle streamwise velocity is reported in figure~\ref{fig:Up}$(a)$. As for the fluid phase, no relevant difference is found in the profiles 
of $U_p(y)$ for the cases studied. Larger variations (also with respect to fluid velocity fluctuations) are found in the profiles of 
$u_{p,rms}', v_{p,rms}', w_{p,rms}'$, depicted in outer units in figures~\ref{fig:Up}$(b),(d),(f)$, and in inner units in 
figures~\ref{fig:Up}$(c),(e),(g)$. From these we can identify two different trends. Very close to the wall (in the viscous sublayer), 
particle velocity fluctuations increase progressively as $\sigma_a/(2a)$ is increased, especially in the streamwise direction. This is probably 
due to the fact that as $\sigma_a/(2a)$ is increased, there are smaller particles that can penetrate more into the viscous and buffer layers. However, 
being smaller and having smaller inertia, they are more easily mixed in all directions due to turbulence structures, and hence experience larger 
velocity fluctuations. Secondly, we observe smaller velocity fluctuations around the centerline for $\sigma_a/(2a)=0.1$. As $\sigma_a/(2a)$ 
increases, larger particles are preferentially found at the centerline and move almost unperturbed in the streamwise direction, hence the reduction in 
$u_{p,rms}', v_{p,rms}', w_{p,rms}'$. Between the viscous sublayer and the centerline, due to turbulence mixing it is difficult to identify an exact 
dependence on $\sigma_a/(2a)$.

\begin{figure}
\centering
\includegraphics[width=.50\textwidth]{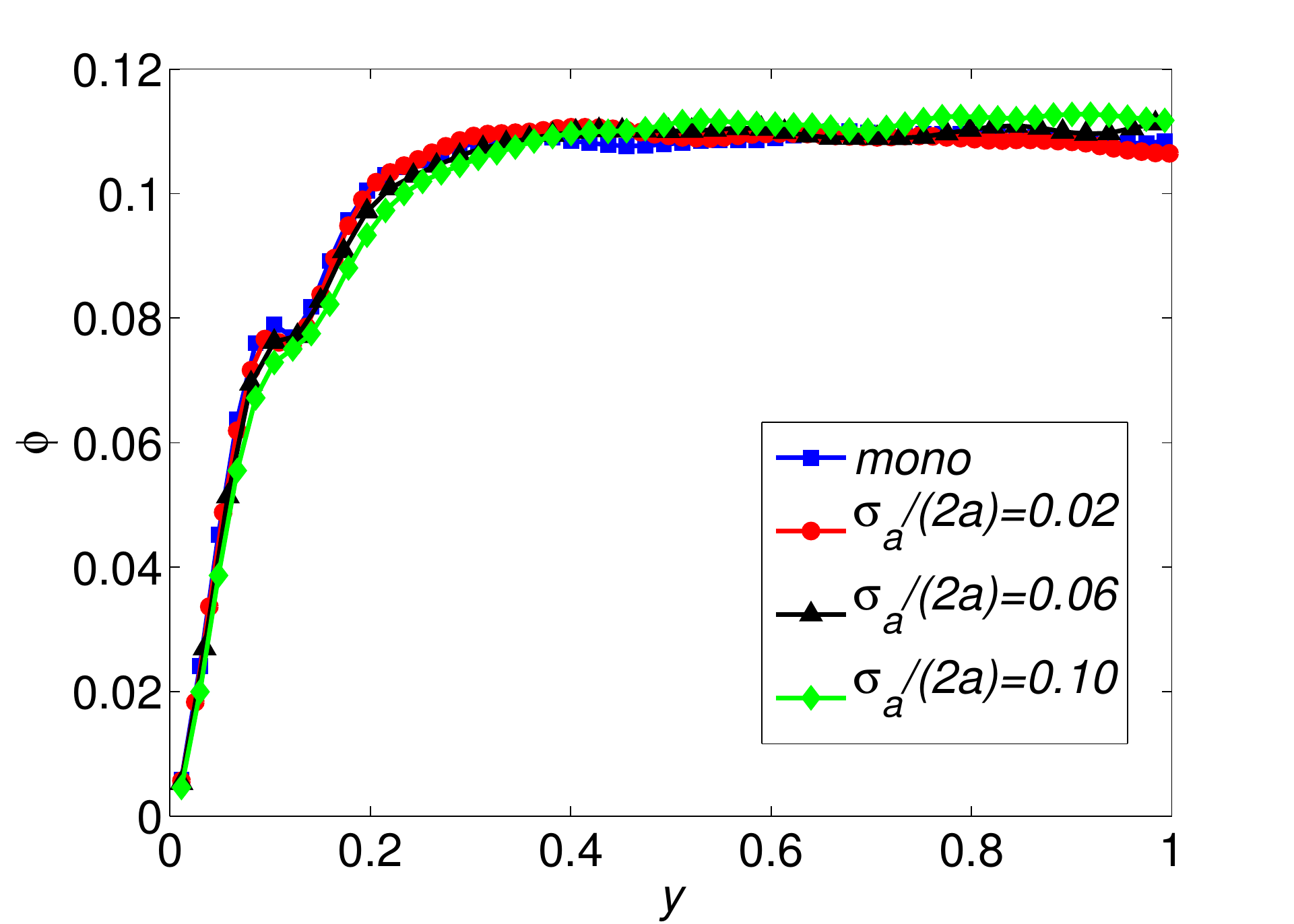}
\put(-185,110){{\large a)}}\\
\includegraphics[width=.50\textwidth]{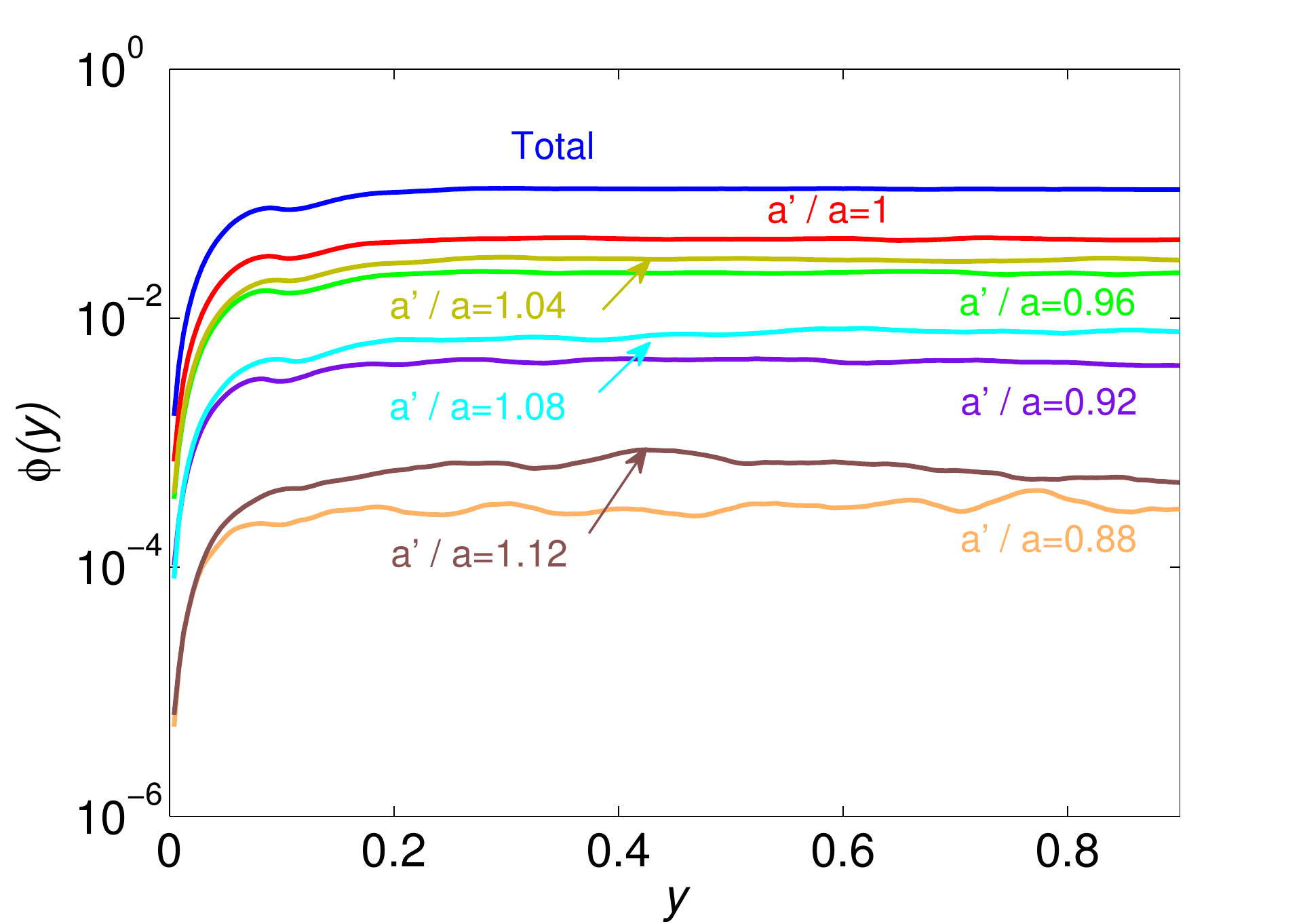}~~~~
\put(-185,110){{\large b)}}
{\includegraphics[width=.50\textwidth]{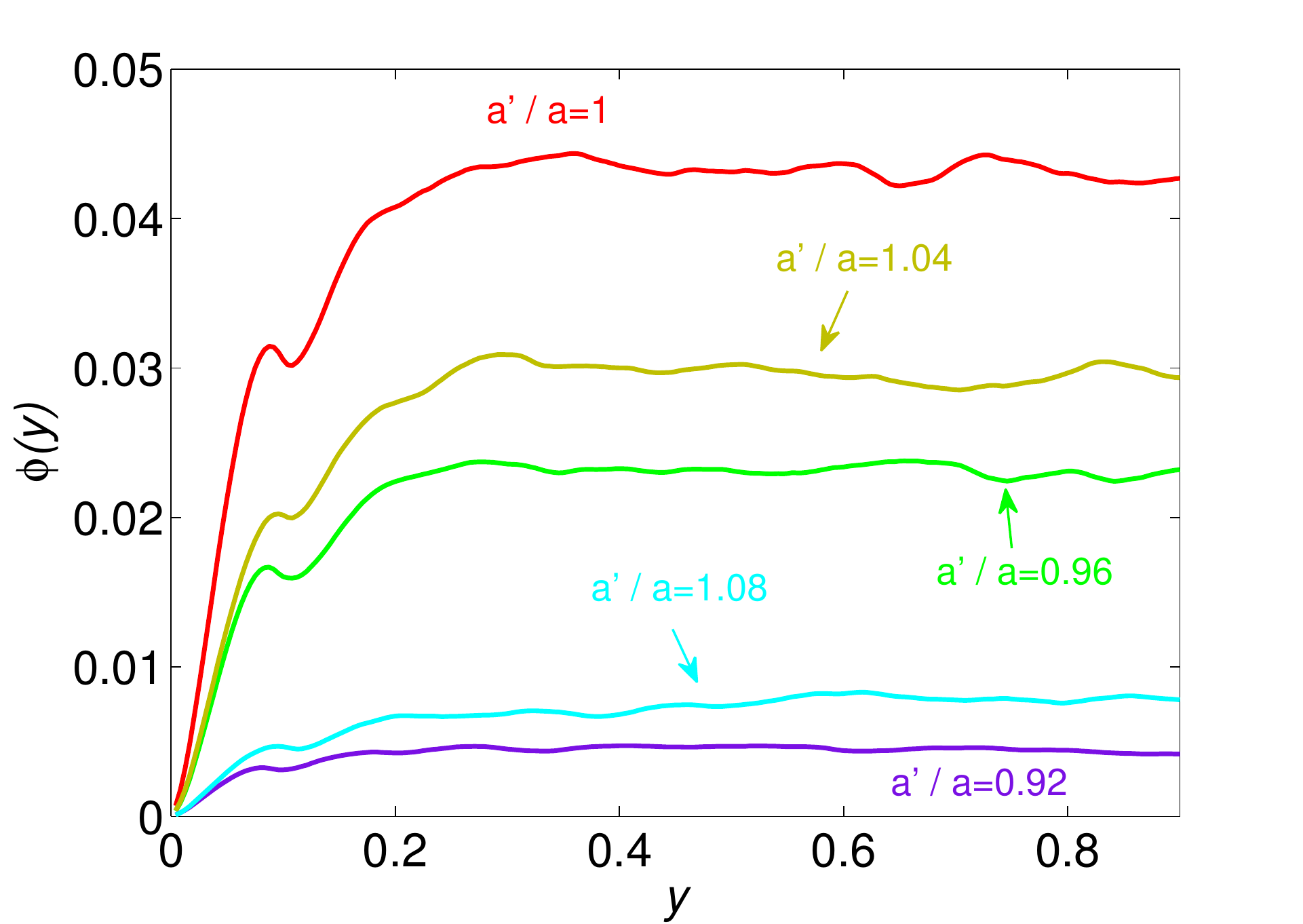}
\put(-180,110){{\large c)}}}
\includegraphics[width=.50\textwidth]{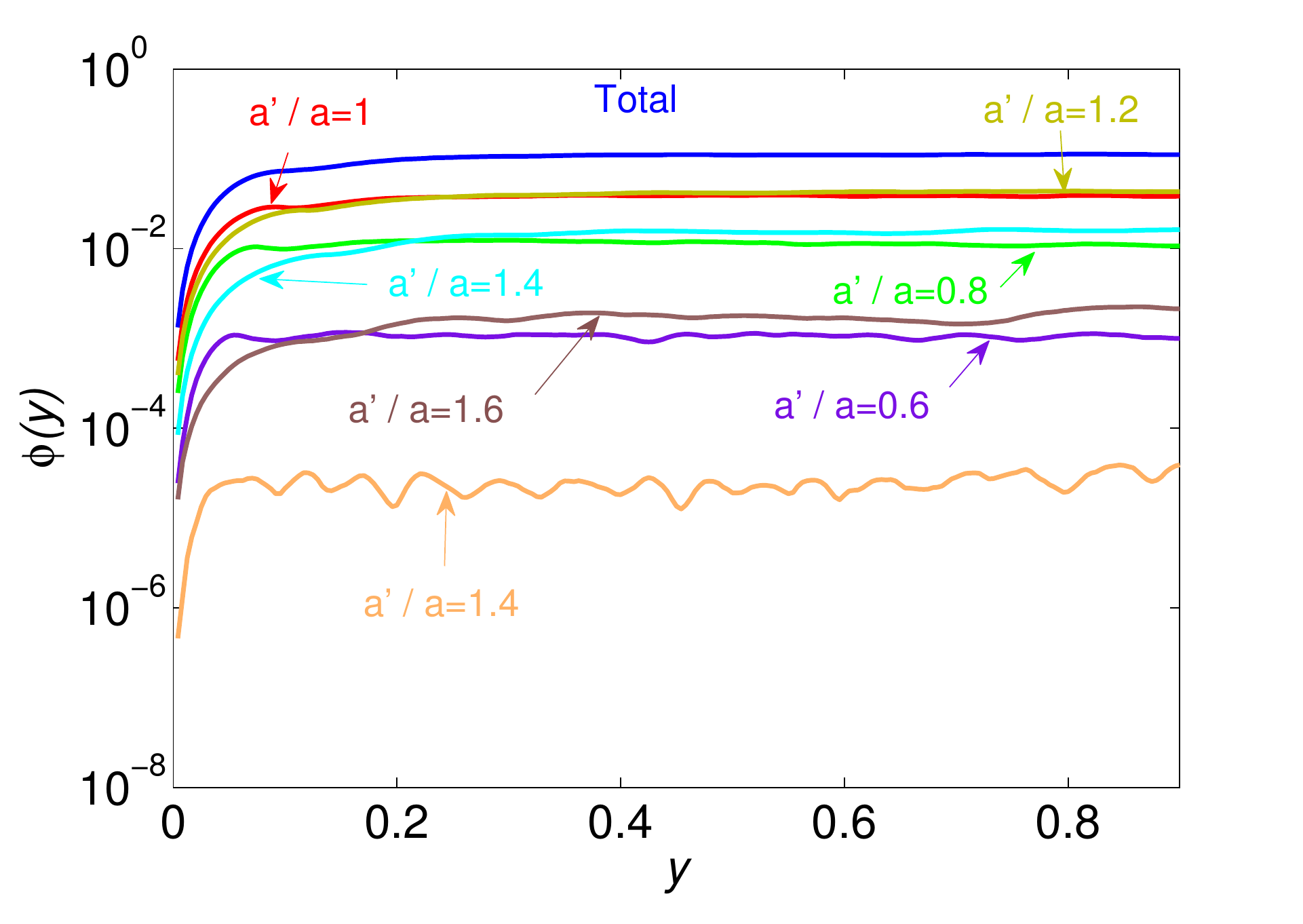}~~~~
\put(-185,110){{\large d)}}
{\includegraphics[width=.50\textwidth]{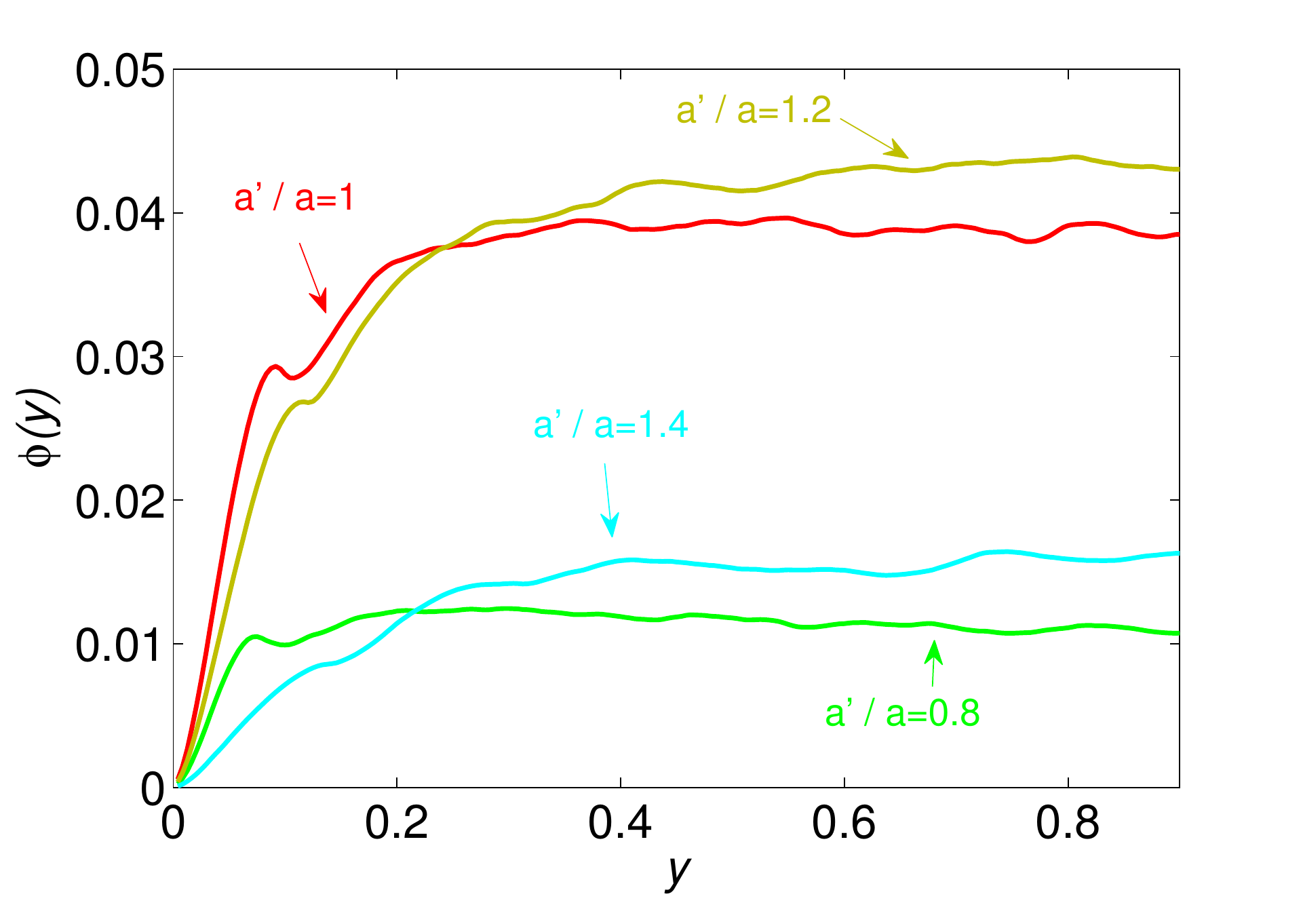}
\put(-180,110){{\large e)}}}
\caption{(a) Mean local volume fraction $\phi(y)$ in the wall-normal direction for different $\sigma_a/(2a)=0$, $0.02$, $0.06$, $0.1$. Mean local volume fraction for 
each particle species of the case with $\sigma_a/(2a)=0.02$: logarithmic (b) and linear (c). Mean local volume fraction for each particle species of the case with 
$\sigma_a/(2a)=0.1$: logarithmic (d) and linear (e).}
\label{fig:phi}
\end{figure}

Concerning the solid phase, we show in figure~\ref{fig:phi} the particle concentration profiles $\phi(y)$ across the channel. From 
figure~\ref{fig:phi}$(a)$ we see that the concentration profiles are similar for all $\sigma_a/(2a)$. However, as previously mentioned we 
notice that as $\sigma_a/(2a)$ is increased, the peak located at $y \simeq 0.1$ is smoothed, while the concentration 
at the centerline is also increased. We then show in figures~\ref{fig:phi}$(b),(d)$ the concentration profiles in logarithmic scale of 
the different species for the cases with $\sigma_a/(2a)=0.02$ and $0.1$; the counterparts in linear scales are shown in 
figures~\ref{fig:phi}$(c),(e)$, where the curves of the species with larger and smaller diameters have been removed for clarity. If we compare 
the different curves to the reference case with $a'/a=1$, we observe that the initial peak moves closer to and further from the walls for 
decreasing and increasing $a'/a$. For larger $a'/a$, the peak is also smoothed until it disappears for $a'/a > 1.2$ in the most extreme case 
with $\sigma_a/(2a)=0.1$. In the latter, for each species with $a'/a > 1$ the concentration grows with $y$ and reaches the maximum 
value at the centerline. On the other hand, the initial peak of the smallest particles is well inside the viscous sublayer. 
 
\begin{figure}
\centering
\includegraphics[width=.50\textwidth]{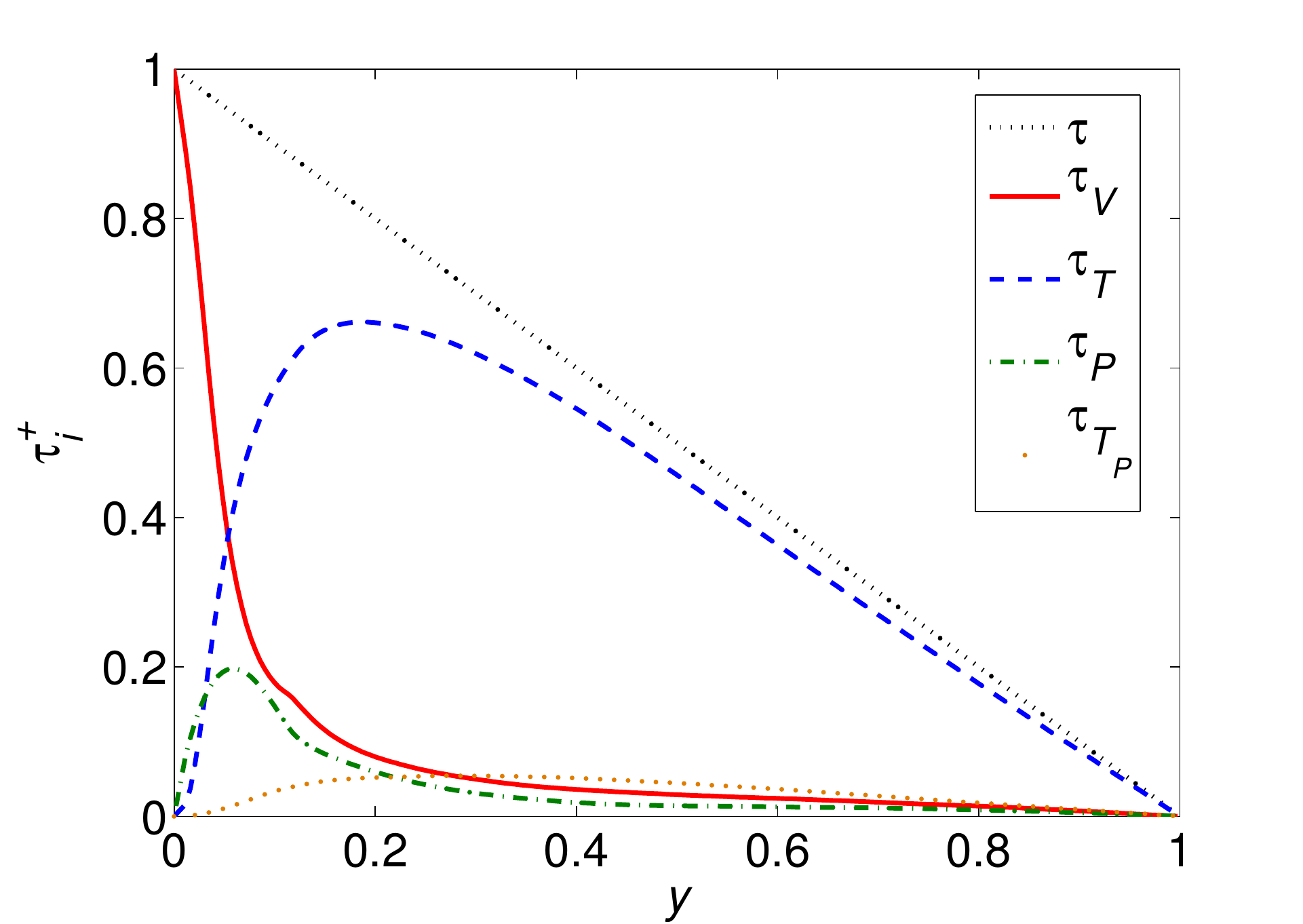}~~~~
\put(-185,110){{\large a)}}
{\includegraphics[width=.50\textwidth]{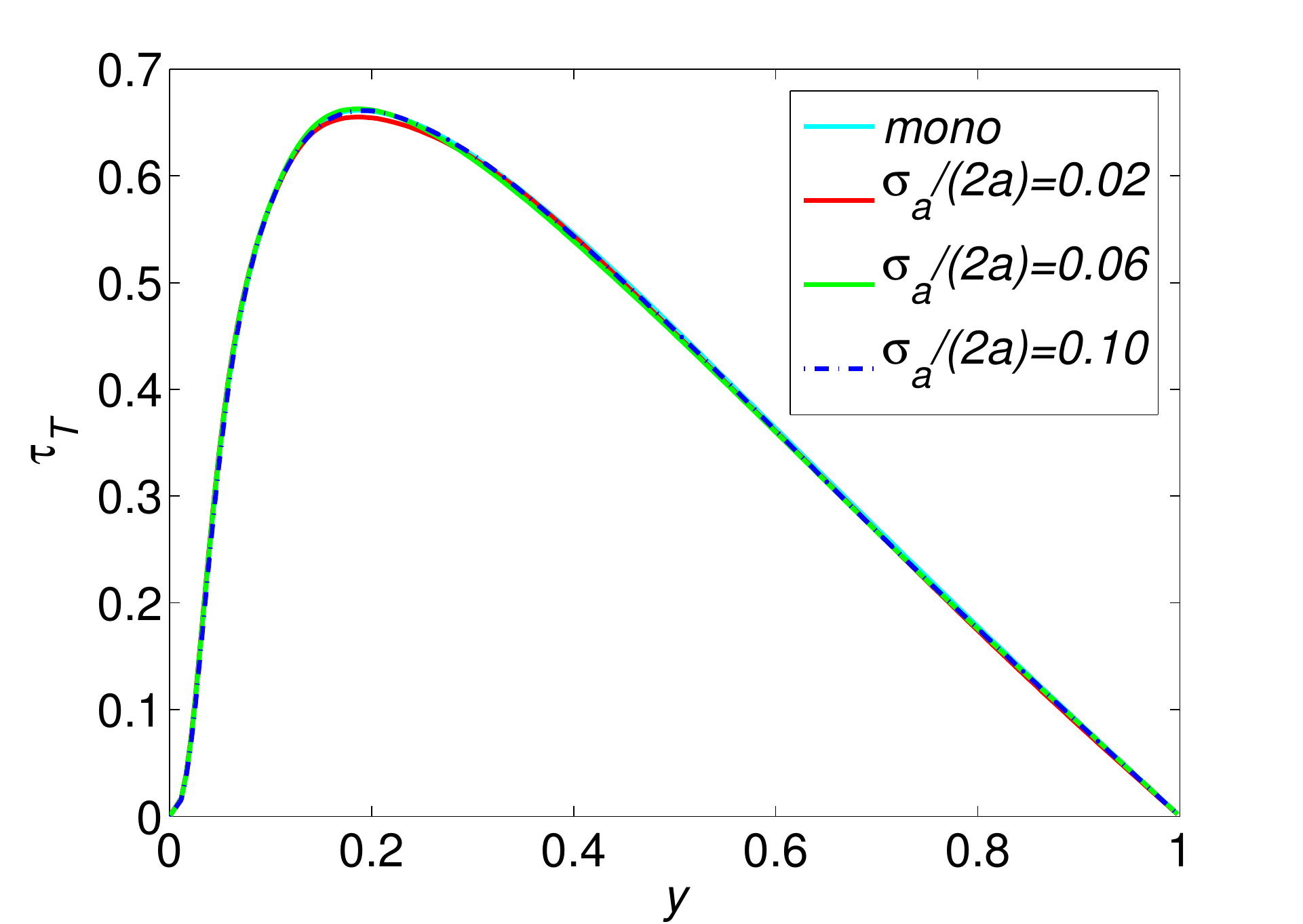}
\put(-180,110){{\large b)}}}
\includegraphics[width=.50\textwidth]{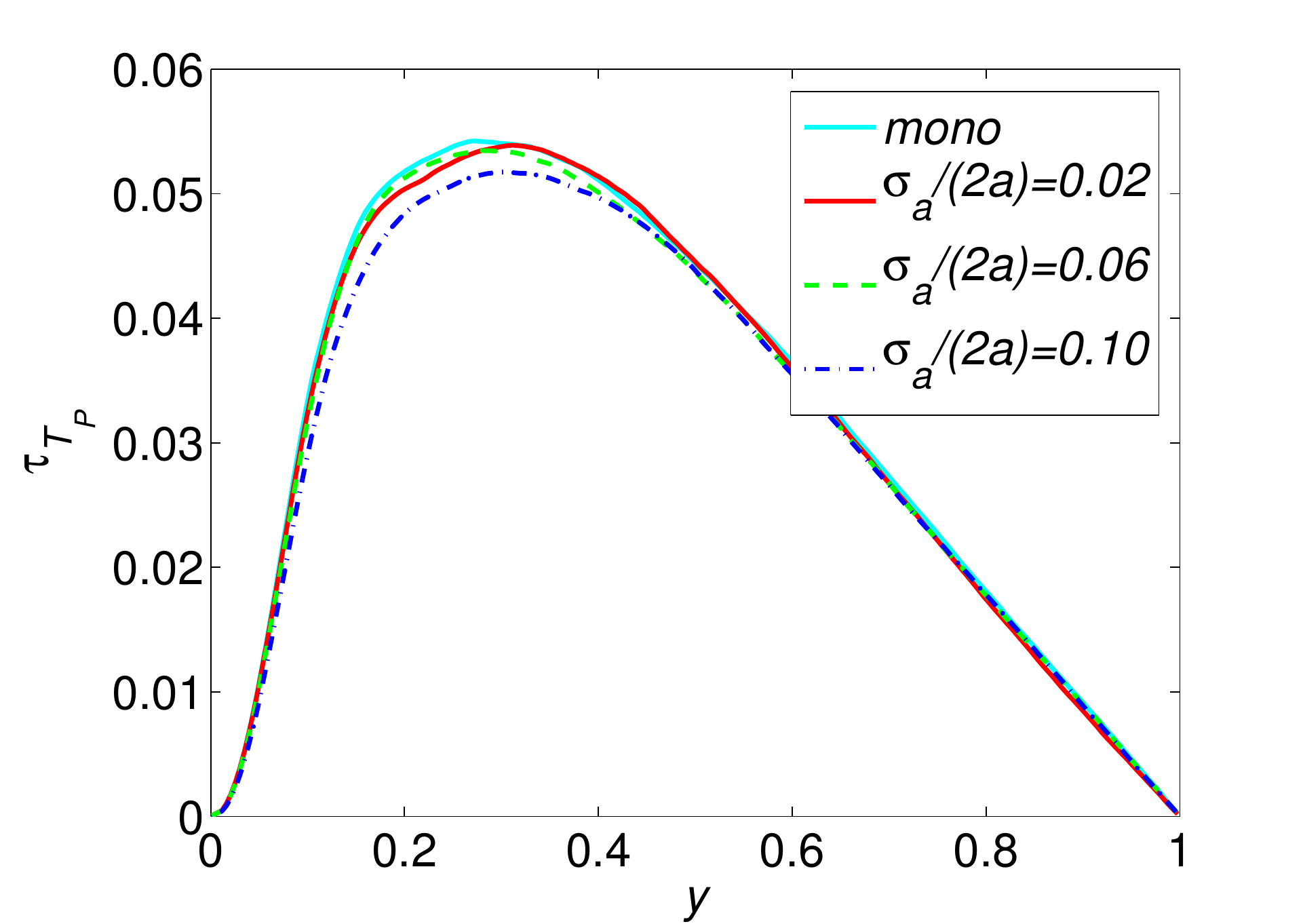}~~~~
\put(-185,110){{\large c)}}
{\includegraphics[width=.50\textwidth]{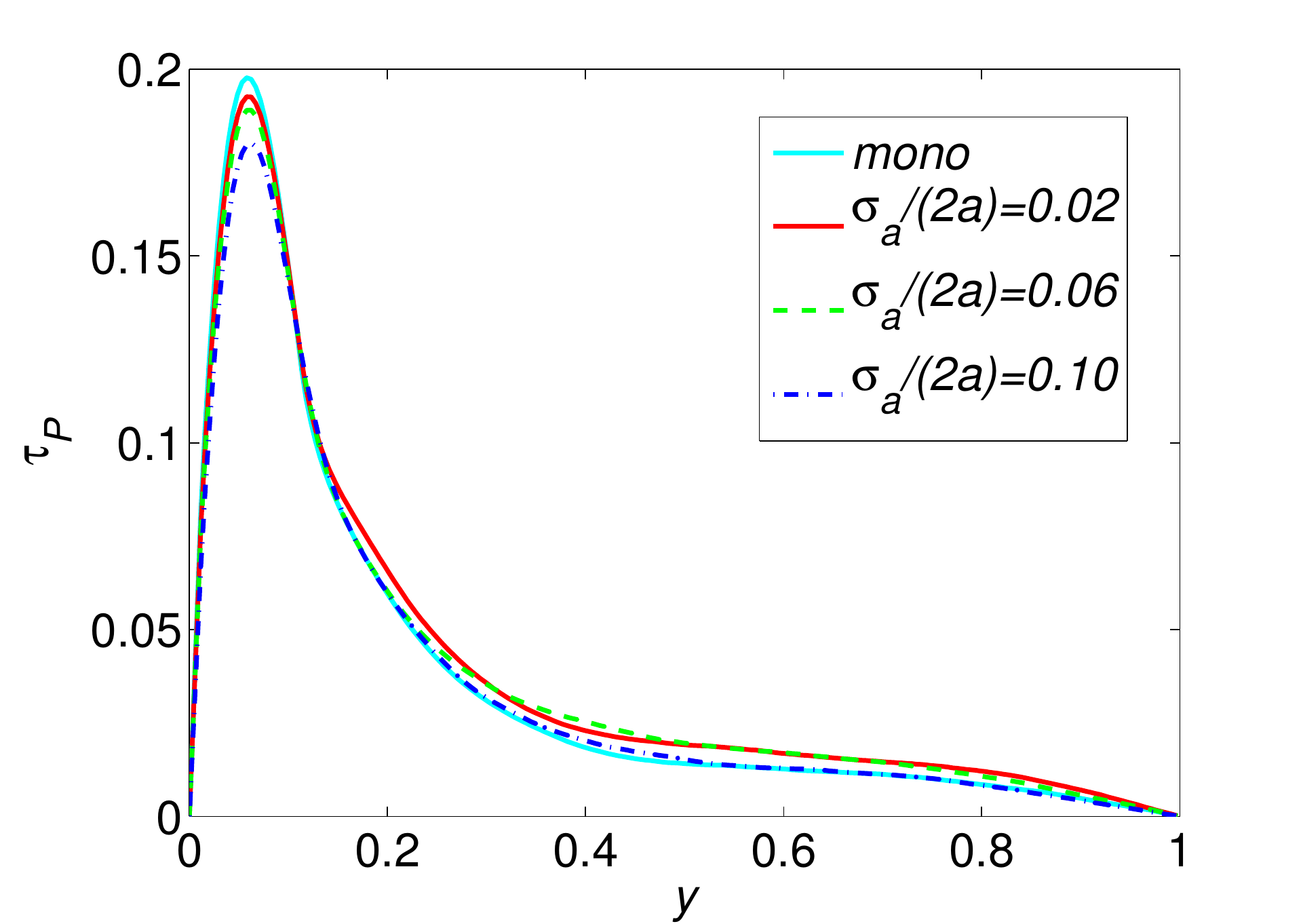}
\put(-180,110){{\large d)}}}
\caption{Shear-stress balance in the wall-normal direction (a). The overall stress $\tau$ is the sum of the viscous stress $\tau_V$, the 
turbulent stres $\tau_T$ (i.e. the Reynolds stress), and the particle-induced stress $\tau_P$; $\tau_{T_P}$ is the particle Reynolds stress. 
Comparison of the turbulent stress (b), particle turbulent stress (c), particle-induced stress (d) for different $\sigma_a/(2a)$.}
\label{fig:stressf}
\end{figure}

We conclude this section by performing a stress analysis. Indeed, the understanding of the momentum exchange between fluid and solid phases 
in particle laden turbulent channel flows is conveniently addressed by examining the streamwise momentum or average stress budget. As in 
Picano et al.\cite{picano2015} we can write the total stress budget (per unit density) as the sum of three terms:
\begin{equation}
\label{stresss}
\tau = \tau_V + \tau_T + \tau_P
\end{equation}
where $\tau = \left. \nu\td{U_{f,x}}{y}\right|_w \left(1-\frac{y}{h}\right)$ is the total stress ($\left. \td{}{y}\right|_w$ denotes a 
derivative taken at the wall), $\tau_V=\nu(1-\phi)\td{U_{f,x}}{y}$ the viscous stress, 
$\tau_T=-\langle u_{c,x}'u_{c,y}'\rangle = -(1-\phi)\langle u_{f,x}'u_{f,y}'\rangle - \phi\langle u_{p,x}'u_{p,y}'\rangle$ 
the turbulent Reynolds shear stress of the combined phase, and $\tau_P= \phi\langle \sigma_{p,xy}/\rho_f \rangle$ the particle induced 
stress. Additionally, we define the particle Reynolds stress $\tau_{T_p}=-\phi\langle u_{p,x}'u_{p,y}'\rangle$. The total stress 
balance for the monodisperse case is shown in figure~\ref{fig:stressf}$(a)$ (the curves for the polydisperse suspensions are not depicted being 
the differences with the actual case negligible). We observe that the major contribution to $\tau$ comes from the turbulent Reynolds stress 
term $\tau_T$ and in particular from the contribution of the fluid phase (the particle Reynolds stress amounts to $\sim 10\%$ of $\tau_T$). 
The particle induced stress $\tau_P$ is important throughout the whole channel (though sub-leading with respect to $\tau_T$) and especially 
close to the wall. In figures~\ref{fig:stressf}$(b),(c),(d)$ we finally compare $\tau_T$, $\tau_{T_p}$ and $\tau_P$ for all $\sigma_a/(2a)$. 
Although the profiles for $\tau_T$ are almost perfectly overlapping, we observe that the maximum of $\tau_{T_p}$ and $\tau_P$ are slightly lower for 
$\sigma_a/(2a)=0.1$. Closer to the centerline $\tau_P$ is smaller for $\sigma/(2a)=0$ and $0.1$.\\ 
Next, we consider the friction Reynolds number $Re_{\tau}=U_*h/\nu$, for each case. For the monodisperse case we have $Re_{\tau}=196$ while 
for the polydisperse cases we obtain $Re_{\tau}=196, 195$ and $194$ for $\sigma_a/(2a)=0.02, 0.06$ and $0.1$. The friction Reynolds number 
is hence larger than that of the unladen case ($Re_{\tau}=180$) due to an enhanced turbulent actvity close to the wall, and to the 
presence of an additional dissipative mechanism introduced by the solid phase (i.e. $\tau_P$)\cite{picano2015,costa2016}. The fact that $Re_{\tau}$ is smaller for 
$\sigma_a/(2a)=0.1$ is related to the fact that the contribution to the total stress from both $\tau_{T_p}$ and $\tau_P$ is slightly reduced 
with respect to all other cases (see figures~\ref{fig:stressf}$(c),(d)$). 
The small discrepancy is however of the order of the statistical error.
 
The results presented clearly show that in turbulent channel flows laden with finite-size spheres, the key parameter in defining both 
rheological properties and turbulence modulation is the solid volume fraction $\phi$. Even in the most extreme case, $\sigma_a/(2a)=0.1$, 
for which the smallest and largest particles have radii of $0.4$ and $1.6$ times that of the reference particles, both fluid and particle 
statistics are similar to those of the monodisperse case. To gain further insight, we also look at the Stokes number of the different 
particles. The Stokes number $St_a$ is the ratio between the typical particle time scale and a characteristic flow time scale. We 
consider the convective time as flow characteristic time $T_f=h/U_0=2h^2/(Re_b\nu)$ and introduce the particle relaxation time defined as 
$T_p = 4Ra^2/(18\nu)$. The effect of finite inertia (i.e. of a non negligible Reynolds number) is taken into account using the correction 
to the particle drag coefficient $C_D$ proposed by Schiller \& Naumann\cite{schil1935}:
\begin{equation}
\label{CD}
C_D = \frac{24}{Re_a} \left(1+0.15Re_a^{0.687}\right)
\end{equation}
Assuming particle acceleration to be balanced only by the nonlinear Stokes drag, and the Reynolds number to be roughly constant, it can 
be found that $V(t) \sim exp\left(-t/T_p'\right)$, where $T_p'=T_p/\left(1+0.15Re_a^{0.687}\right)$. 
For sake of simplicity and in first approximation we define a shear-rate based particle Reynolds number $Re_a=Re_b(a/h)^2$. The final 
expression for the modified Stokes number is
\begin{equation}
\label{Stokes}
St_a' = \frac{T_p}{T_f} \frac{1}{\left(1+0.15Re_a^{0.687}\right)}= \left(\frac{2a}{h}\right)^2 \frac{1}{36} Re_b R \frac{1}{\left(1+0.15Re_a^{0.687}\right)}
\end{equation}
For the reference particles we obtain $Re_a = 17.3$ and $St_a'=0.93$. For the smallest particles ($a'/a=0.4$) we find $Re_a=2.8$ and 
$St_a'=0.24$, while for the largest ($a'/a=1.6$) $Re_a=44.2$ and $St_a'=1.63$. Hence, when the radius of the largest particles is $4$ times 
that of the smallest particles, there is an order of magnitude difference in the Stokes number. It is also interesting to note 
that albeit the use of a nonlinear drag correction, if we average the Stokes numbers of largest and smallest particles we get that of the 
reference case ($St_a'=0.93$). Hence, $30\%$ of the particles respond more slowly to fluid-induced velocity perturbations than the reference 
particles, while other $30\%$ respond more quickly. On average, however, the suspension responds with a time scale comparable to that of the 
monodisperse case, therefore behaving similarly from a statistical perspective. We expect this to be the case for all volume fractions 
$\phi$ in this semi-dilute regime. This finding can be useful for modeling the behavior of rigid-particle suspensions.

\begin{figure}
\centering
\includegraphics[width=.50\textwidth]{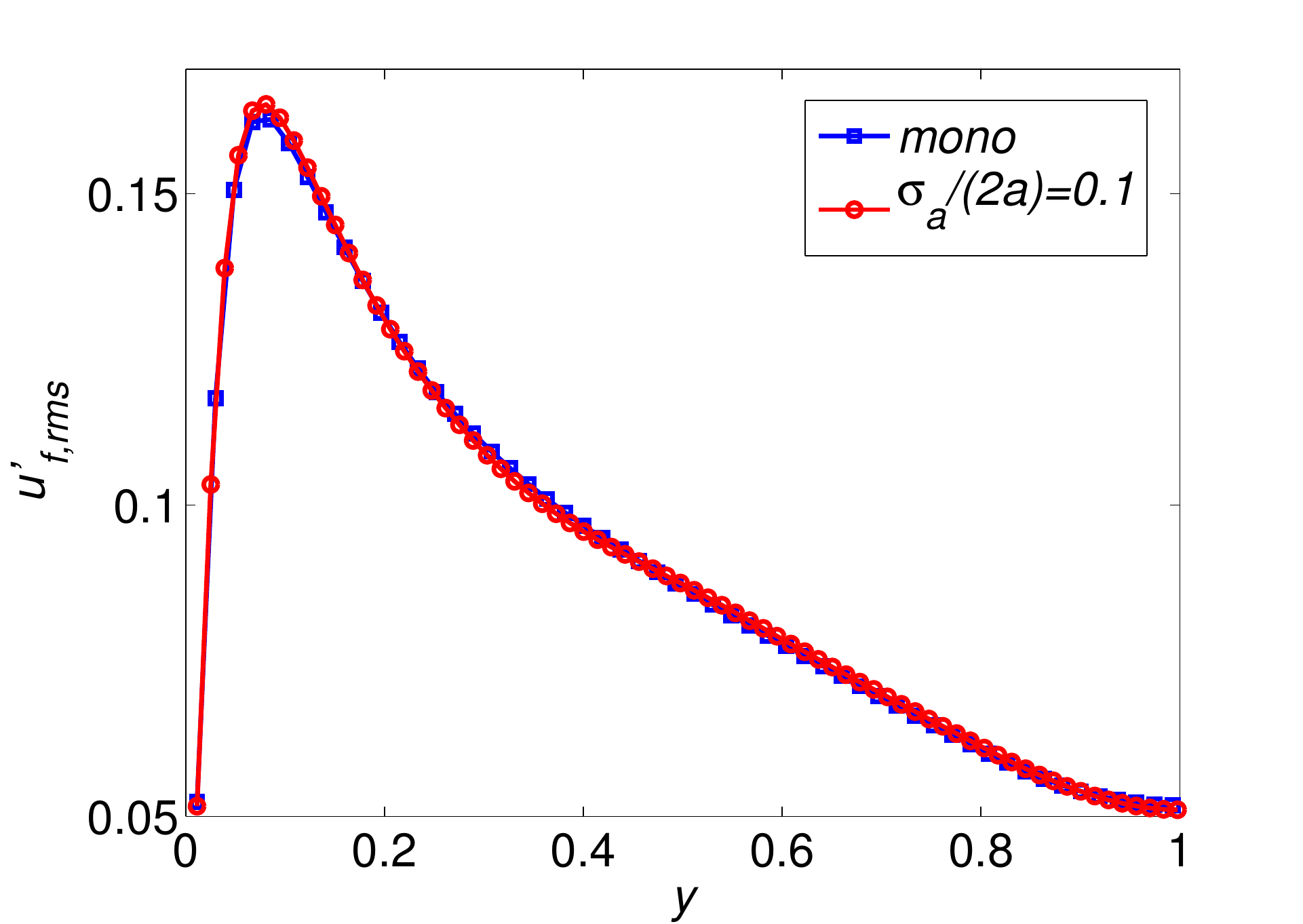}~~~~
\put(-185,110){{\large a)}}
{\includegraphics[width=.50\textwidth]{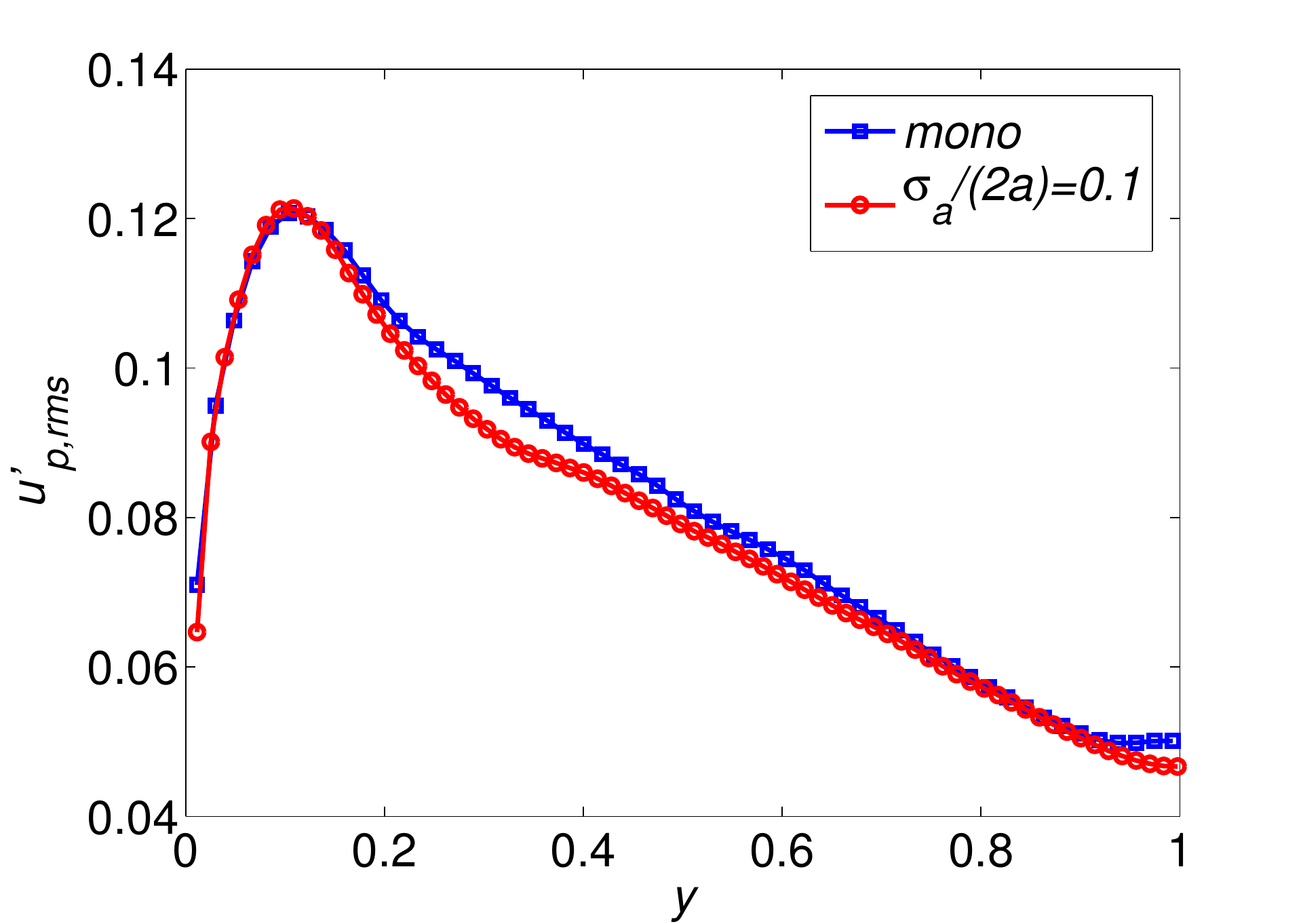}
\put(-180,110){{\large b)}}}
\includegraphics[width=.50\textwidth]{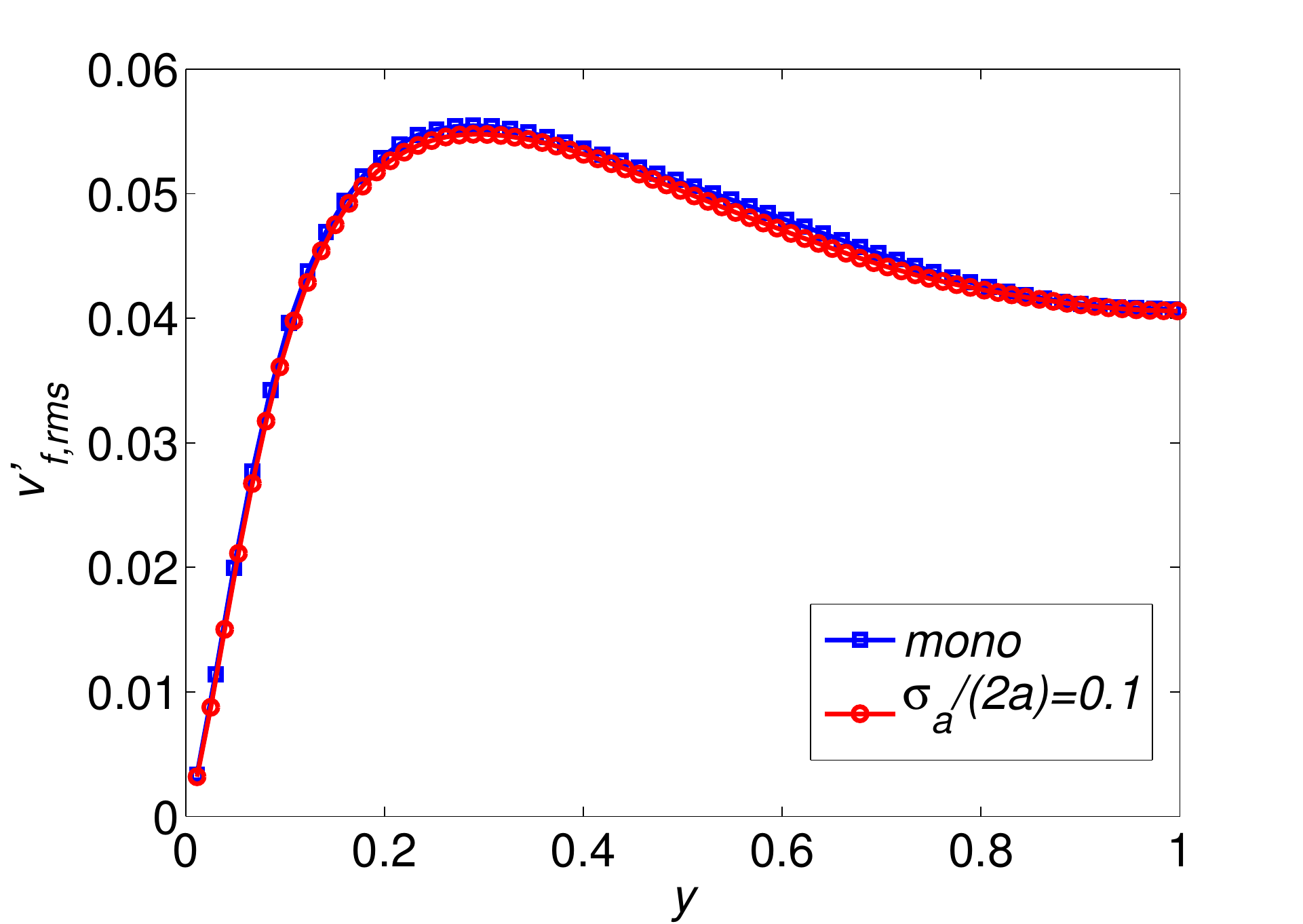}~~~~
\put(-185,110){{\large c)}}
{\includegraphics[width=.50\textwidth]{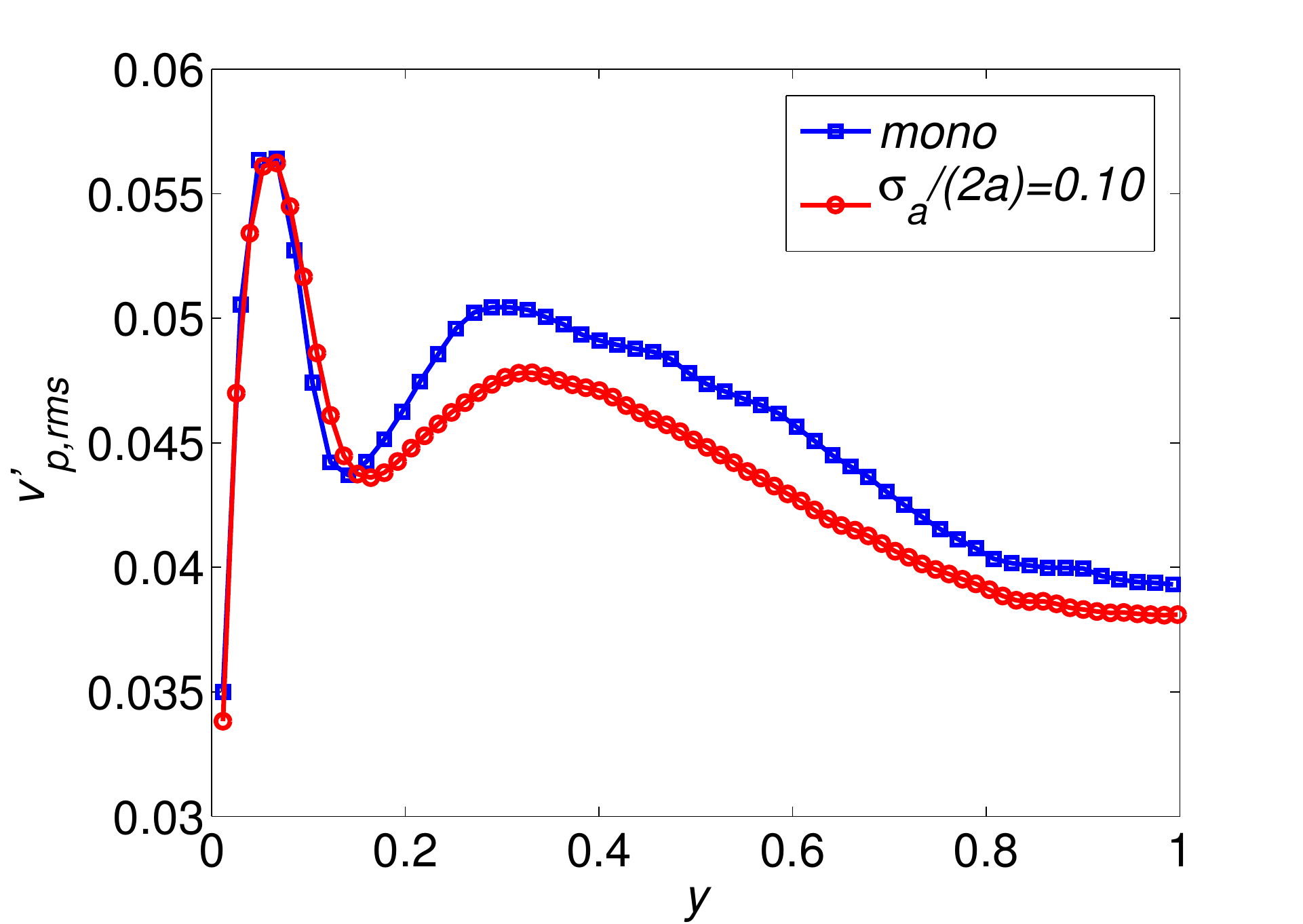}
\put(-180,110){{\large d)}}}
\includegraphics[width=.50\textwidth]{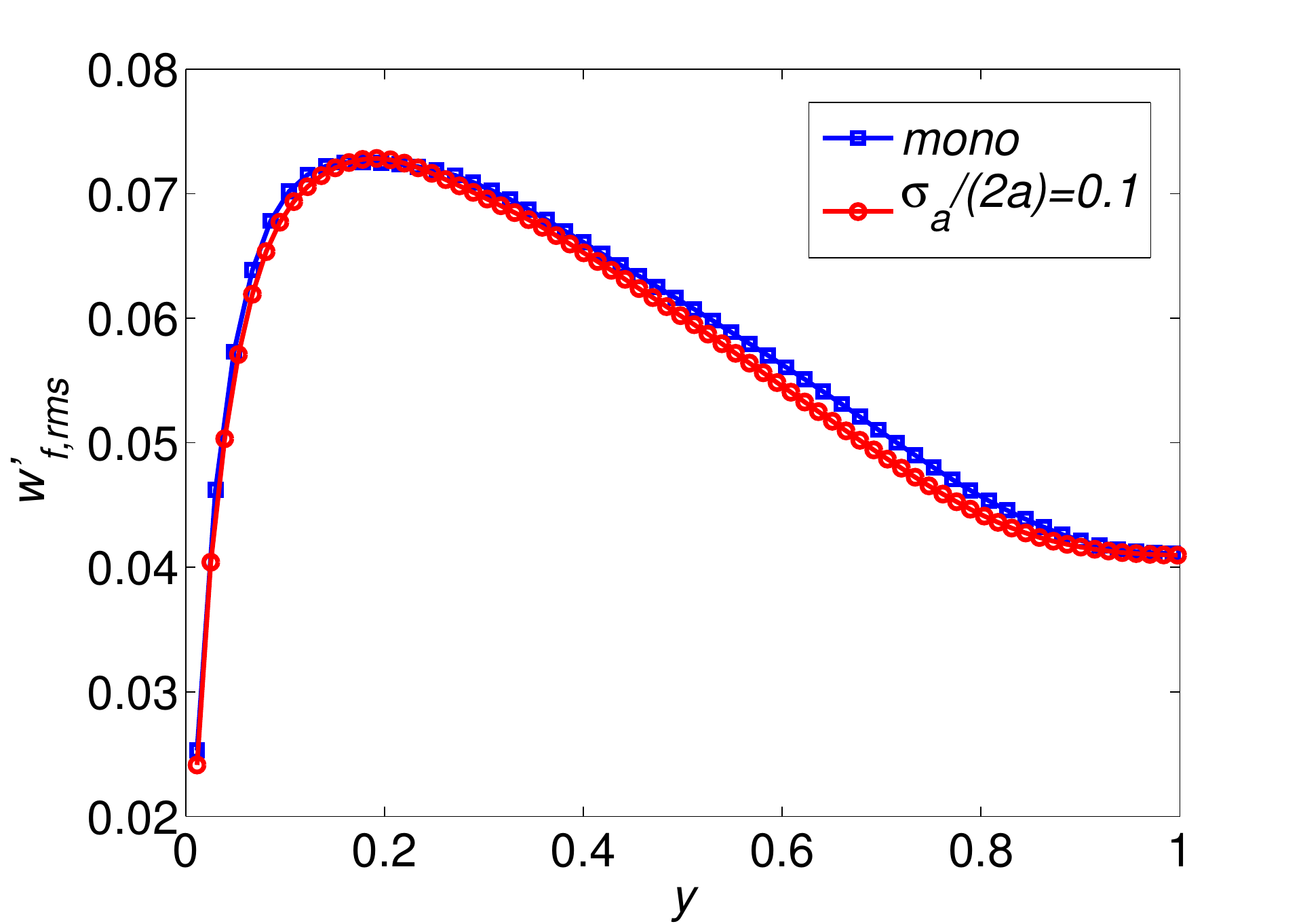}~~~~
\put(-185,110){{\large e)}}
{\includegraphics[width=.50\textwidth]{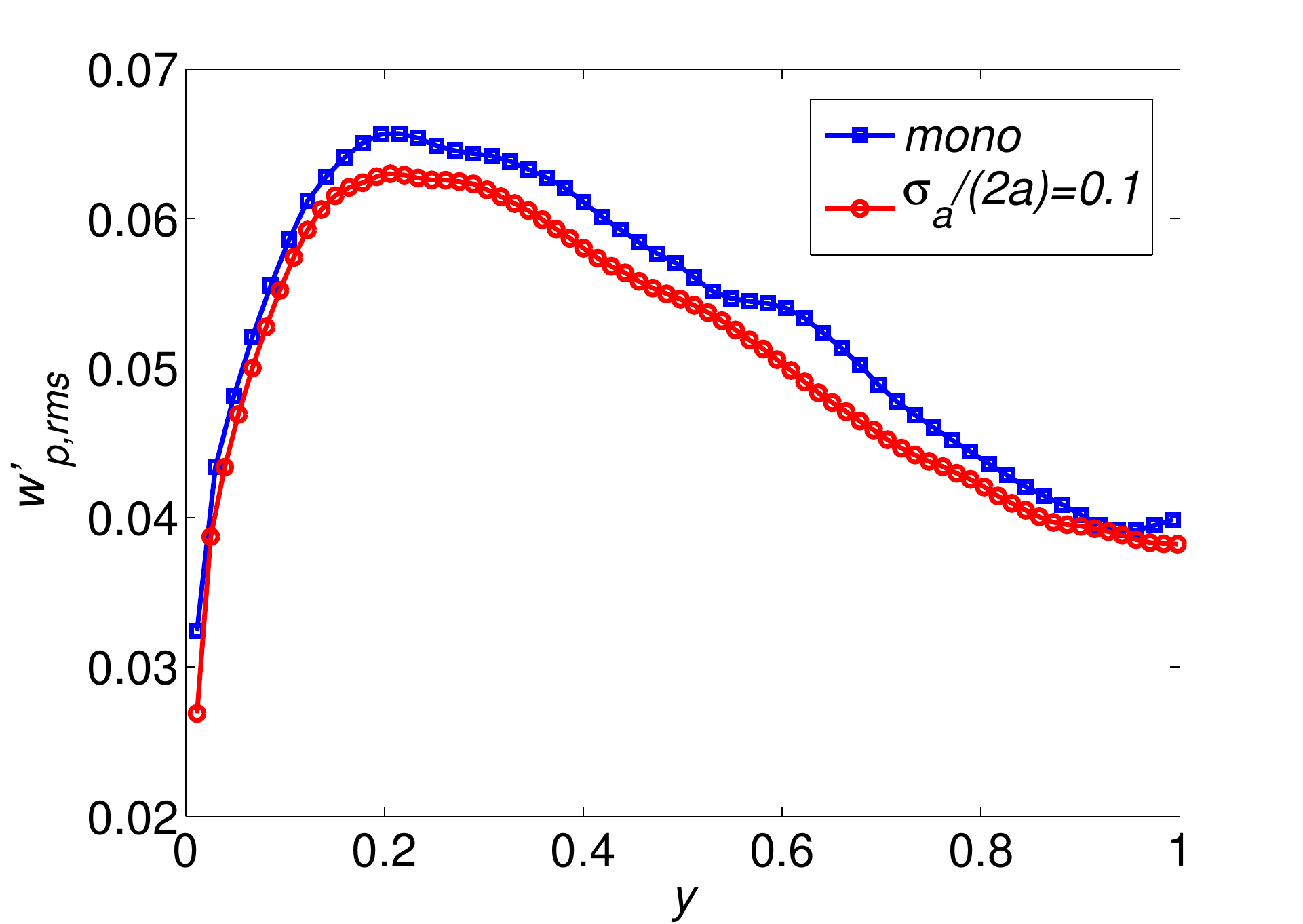}
\put(-180,110){{\large f)}}}
\caption{Fluid and particle velocity fluctuations in outer units, in the streamwise (a)-(b), 
wall-normal (c)-(d) and spanwise (e)-(f) directions, for $\sigma_a/(2a)=0$ and $0.1$.}
\label{fig:Up2}
\end{figure}

To check this, we performed 2 additional simulations with $\phi=2\%$ and $\sigma_a/(2a)=0$ and $0.1$. The fluid and particle velocity 
fluctuations in the streamwise, wall-normal and spanwise directions are shown in figures~\ref{fig:Up2}$(a),(c),(e)$ and in 
figures~\ref{fig:Up2}$(b),(d),(f)$. The mean fluid and particle streamwise velocities are not reported since the 
curves are again almost perfectly overlapping. Regarding the fluid velocity fluctuation profiles, we see that the results of the mono and 
polydisperse cases are almost identical. As for $\phi=10\%$, particle velocity fluctuation profiles exhibit larger variations with respect 
to the monodisperse results. In particular, we notice that the profiles vary in a similar way for both $\phi$: smaller fluctuations 
throughout the channel, except in the viscous sublayer where the maxima of streamwise and wall-normal fluctuations 
are found ($y \in (0.1;0.2)$). However, the largest relative difference between the velocity fluctuation profiles of the mono and polydisperse 
cases is only about $7\%$.\\
Finally, we also computed the friction Reynolds number and found a similar behavior as for $\phi=10\%$. Indeed, for both 
$\phi$ and $\sigma_a/(2a)=0.1$, the friction Reynolds number $Re_{\tau}$ decreases by about $1\%$ with respect to the case with 
$\sigma_a/(2a)=0$. For $\phi=2\%$, $Re_{\tau}$ decreases from $186$ to $183$.
%\begin{figure}
%\centering
%\includegraphics[width=.50\textwidth]{FIG/ufrms2.eps}~~~~
%\put(-185,110){{\large a)}}
%{\includegraphics[width=.50\textwidth]{FIG/uprms2.eps}
%\put(-180,110){{\large b)}}}
%\includegraphics[width=.50\textwidth]{FIG/vfrms2.eps}~~~~
%\put(-185,110){{\large c)}}
%{\includegraphics[width=.50\textwidth]{FIG/vprms2.eps}
%\put(-180,110){{\large d)}}}
%\includegraphics[width=.50\textwidth]{FIG/wfrms2.eps}~~~~
%\put(-185,110){{\large e)}}
%{\includegraphics[width=.50\textwidth]{FIG/wprms2.eps}
%\put(-180,110){{\large f)}}}
%\caption{Fluid and particle velocity fluctuations in outer units, in the streamwise (a)-(b), 
%wall-normal (c)-(d) and spanwise (e)-(f) directions, for $\sigma_a/(2a)=0$ and $0.1$.}
%\label{fig:Up2}
%\end{figure}

\subsection{Single-point particle statistics}

\begin{figure}
\centering
\includegraphics[width=.50\textwidth]{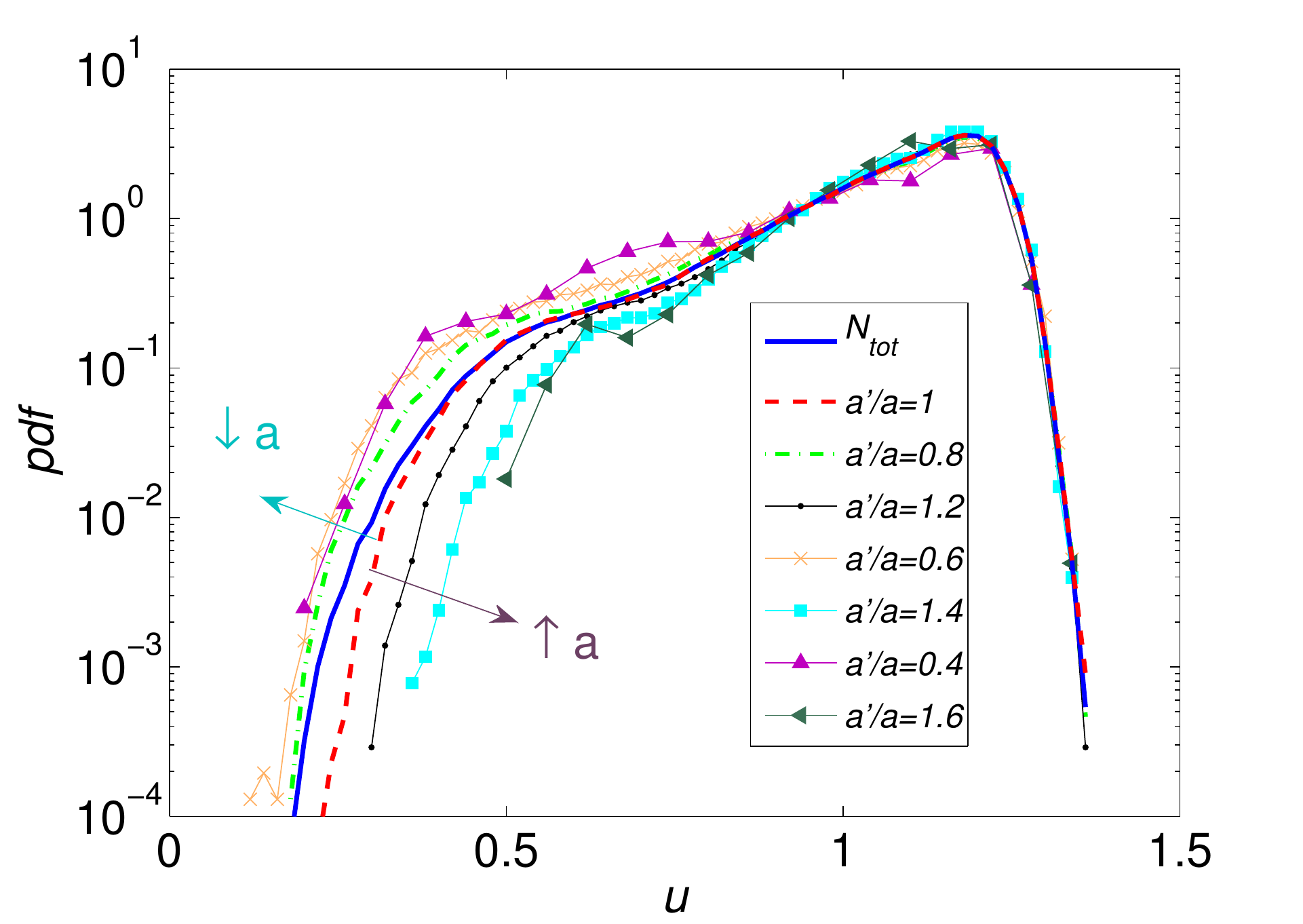}~~~~
\put(-185,110){{\large a)}}
{\includegraphics[width=.50\textwidth]{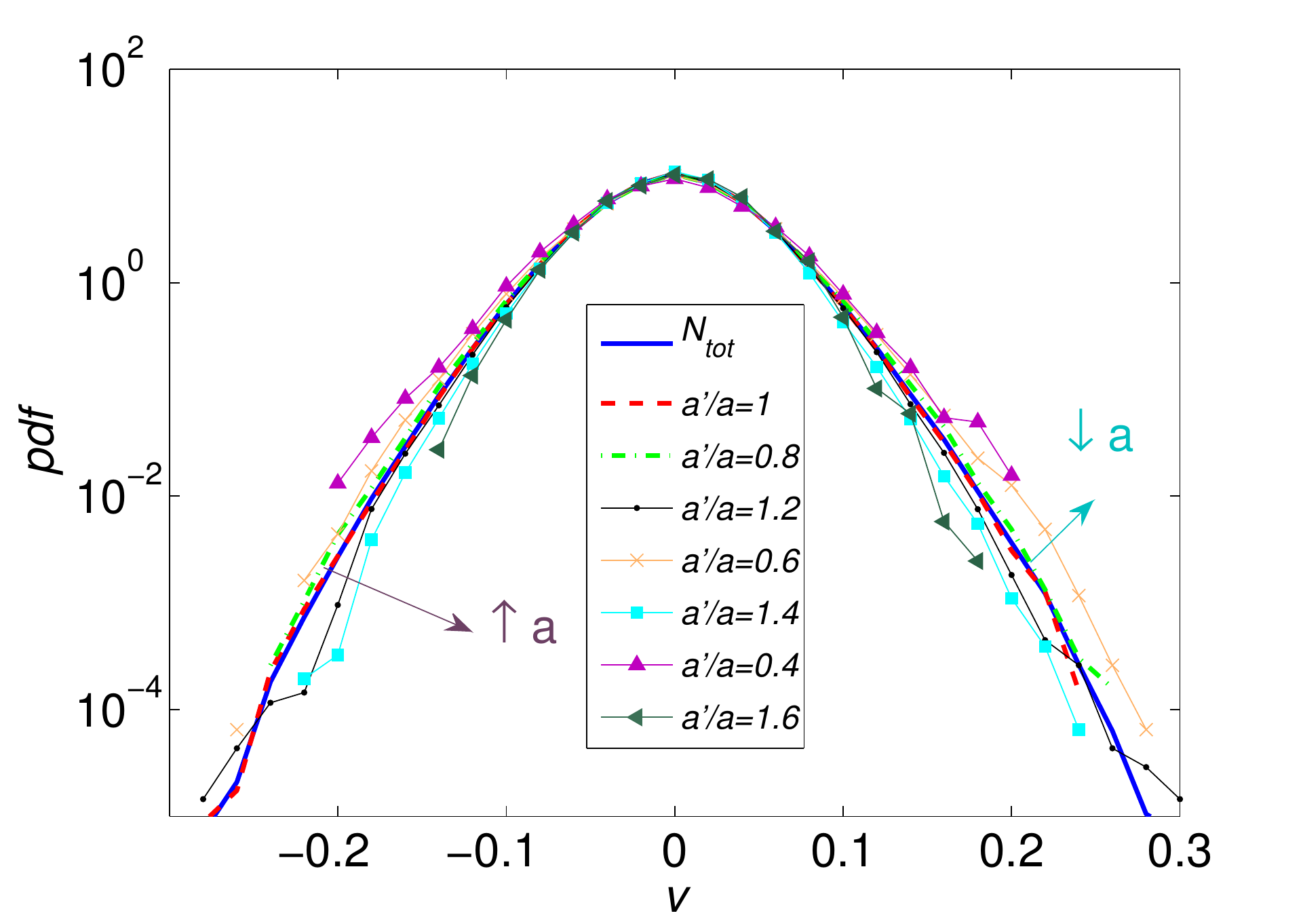}
\put(-185,110){{\large b)}}}
\caption{Probability density functions of particle streamwise (a) and wall-normal velocities (b), 
for $\sigma_a/(2a)=0.1$.}
\label{fig:pdf}
\end{figure}

We wish to give further insight on the behavior of the solid phase dynamics by examining the probability density functions, $pdf$s, of 
particle velocities. In particular, we report the results obtained for the polydisperse suspension with $\sigma_a/(2a)=0.1$, as this 
revealed to be the most interesting case in the previous section. The distributions of the streamwise and wall-normal components of 
the particle velocity are calculated in the whole channel (for each particle species) and are depicted in figures~\ref{fig:pdf}$(a)$ 
and $(b)$. The $pdf$ of the spanwise component is not shown since it is qualitatively similar to the wall-normal one. For both 
components, the $pdf$s of particles with different radius $a'$ are similar around the modal value. The larger differences are found 
in the tails of the $pdf$s and hence we report them in logarithmic scale.

Concerning the $pdf$s of streamwise particle velocities, we see that the variance $\sigma_{u}^2$ increases as the particle radius is 
reduced, while it decreases for increasing $a'$. In particular, the $pdf$s are identical for velocities higher than the modal value while 
the larger differences are found in the low velocity tails. Smaller particles are indeed able to closely approach the walls and hence 
translate with lower velocities than larger particles. Having in mind the profile of the mean streamwise velocity in a channel 
flow, it is then clear that larger particles whose centroids are more distant from the walls, translate more quickly than smaller 
particles.\\
The $pdf$s of the wall-normal velocities show less differences when varying $a'$. The variance is similar for all species. One can 
however still notice that the variance slightly increases for smaller particles (smaller $a'$) while it decreases for larger ones (
larger $a'$). As discussed in the previous section, smaller particles have smaller Stokes numbers (i.e. smaller inertia) and are 
perturbed more easily by turbulence structures thereby reaching higher velocities (with higher probability) than larger particles.

\begin{figure}
\centering
\includegraphics[width=.50\textwidth]{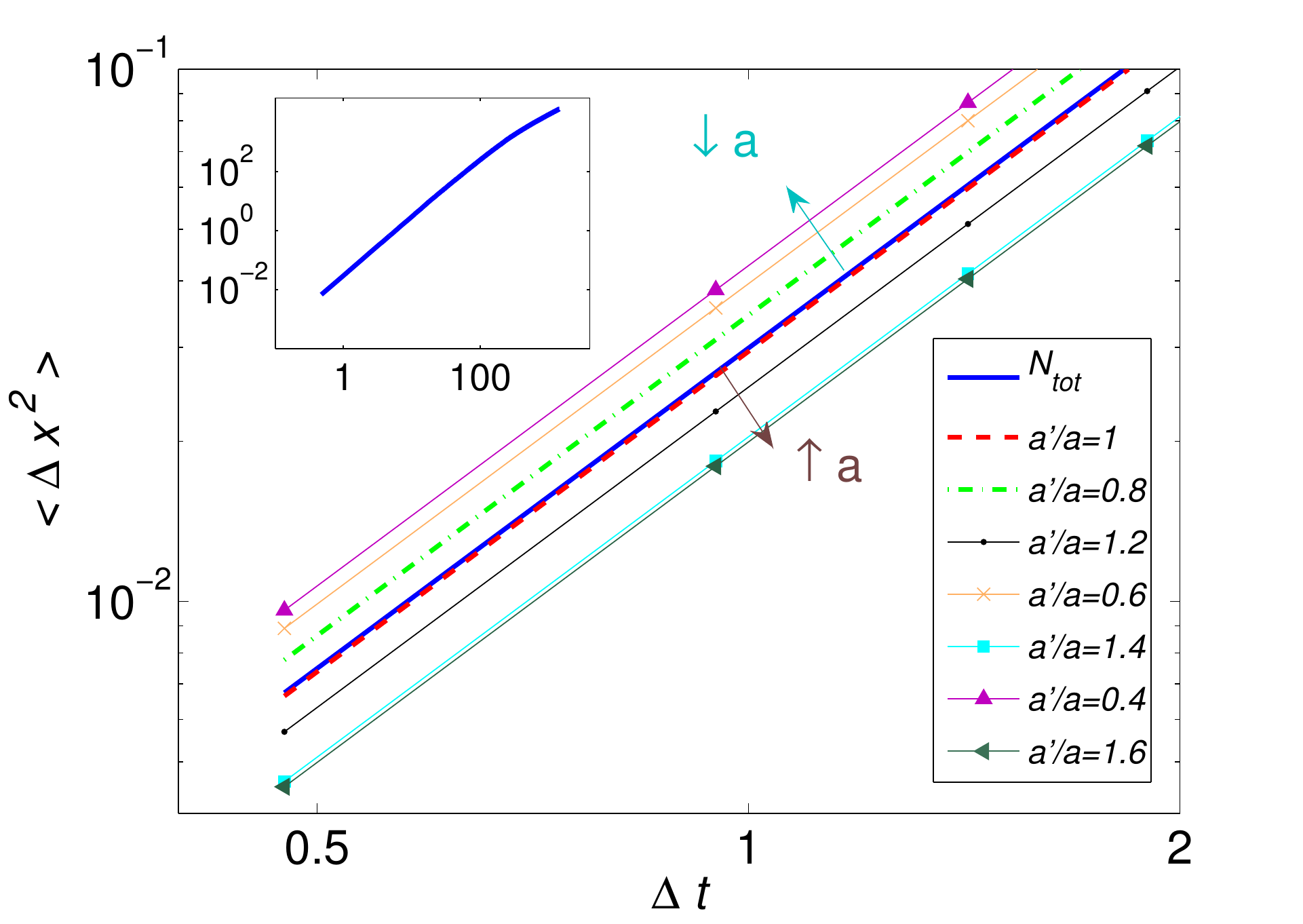}~~~~
\put(-185,110){{\large a)}}
{\includegraphics[width=.50\textwidth]{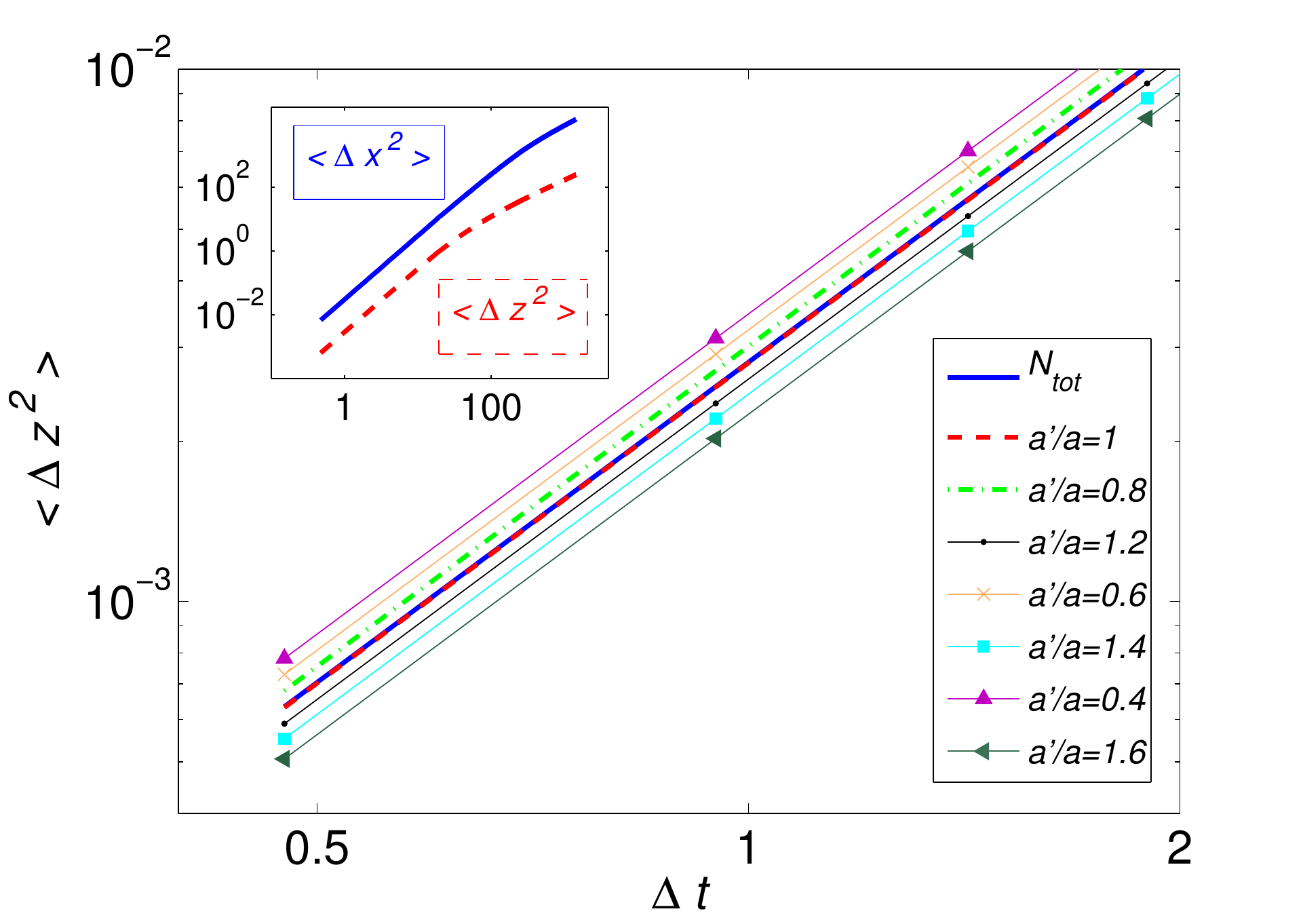}
\put(-180,110){{\large b)}}}
\caption{Mean-squared displacement of particles in the streamwise (a) and spanwise directions (b) for $\sigma_a/(2a)=0.1$.}
\label{fig:msd}
\end{figure}

%\begin{figure}
%\centering
%\includegraphics[width=.50\textwidth]{FIG/dispy7d.eps}~~~~
%\put(-185,110){{\large a)}}
%{\includegraphics[width=.50\textwidth]{FIG/dispx7d.eps}
%\put(-180,110){{\large b)}}}
%\caption{Mean-squared displacement of particles in the streamwise (a) and spanwise directions (b) for $\sigma_a/(2a)=0.1$.}
%\label{fig:msd}
%\end{figure}
%
Finally, we discuss the particles dispersion in the streamwise and spanwise directions. Particle motion is constrained 
in the wall-normal direction by the presence of the walls and is therefore not examined here. The dispersion is quantified as the 
variance of the particle displacement as function of the separation time $\Delta t$ (i.e. the mean-square displacement of particle 
trajectories)
\begin{equation}
\label{disp}
\langle \Delta \vec x^2 \rangle(\Delta t) = \langle \left[\vec x_p(\bar t +  \Delta t) - \vec x_p(\bar t) \right]^2 \rangle_{p,\bar t}
\end{equation}
where $\langle \cdot \rangle_{p,\bar t}$ denotes averaging over time $\bar t$ and the number of particles $p$.\\
The mean-square displacement in the streamwise direction is shown in figure~\ref{fig:msd}$(a)$. From the subplot we see that initially, 
in the so-called ballistic regime, particle dispersion $\langle \Delta x^2 \rangle$ shows a quadratic dependence on time. Only after 
$\Delta t \sim 100 (2a)/U_0$ the curve starts to approach the linear behavior typical of a diffusive motion. As expected, we observe 
that smaller particles have a larger mean-square displacement than larger particles in the ballistic regime. 
However, the difference between $\langle \Delta \vec x^2 \rangle$ for the smallest and largest particles ($a'/a=0.4$ and $1.6$) is limited.

Concerning the dispersion in the spanwise direction (figure~\ref{fig:msd}$(b)$), we clearly notice that $\langle \Delta z^2 \rangle$ 
is $1$ and $2$ orders of magnitude smaller than $\langle \Delta x^2 \rangle$ in the ballistic and diffusive regimes. The latter is also 
reached earlier than in the streamwise direction, due to the absence of a mean flow. The discussion of the previous paragraph on the effect 
of particle size on dispersion in the ballistic regime, applies also in the present case. 
However, as the diffusive regime is approached, the mean-squared displacements $\langle \Delta z^2 \rangle$ of all $a'/a$ become more 
similar. For each $a'/a$ we also find that the diffusion coeffient, defined as $D_{p,z} \simeq \langle \Delta z^2 \rangle/(2\Delta t)$, is 
approximately $0.08$. A remarkable and not yet understood difference is found for $a'/a=1.4$, for which $D_{p,z}$ is found to be $7\%$ larger. 
This could be 
due to the fact that less samples are used to calculate $\langle \Delta z^2 \rangle$.
The number of particles with $a'/a=1.4$ is indeed substantially smaller than for $a'/a \in [0.8;1.2]$. 

To conclude this section, we emphasize that particle related statistics (probability density functions of velocities and 
mean-square displacements) only slightly vary for different $a'/a$. In particular, the $pdf$s of particle velocities for smaller particles are 
wider than those of the larger particles. Accordingly, the mean-squared displacement $\langle \Delta \vec x^2 \rangle$ of particles with $a'/a < 1$ 
is larger than that for particles with $a'/a > 1$, at least in the ballistic regime. Indeed, in the spanwise direction we find that the 
diffusion coeffients are approximately similar for all species.

\subsubsection{Particle collision rates}

\begin{figure}
\centering
\includegraphics[width=.50\textwidth]{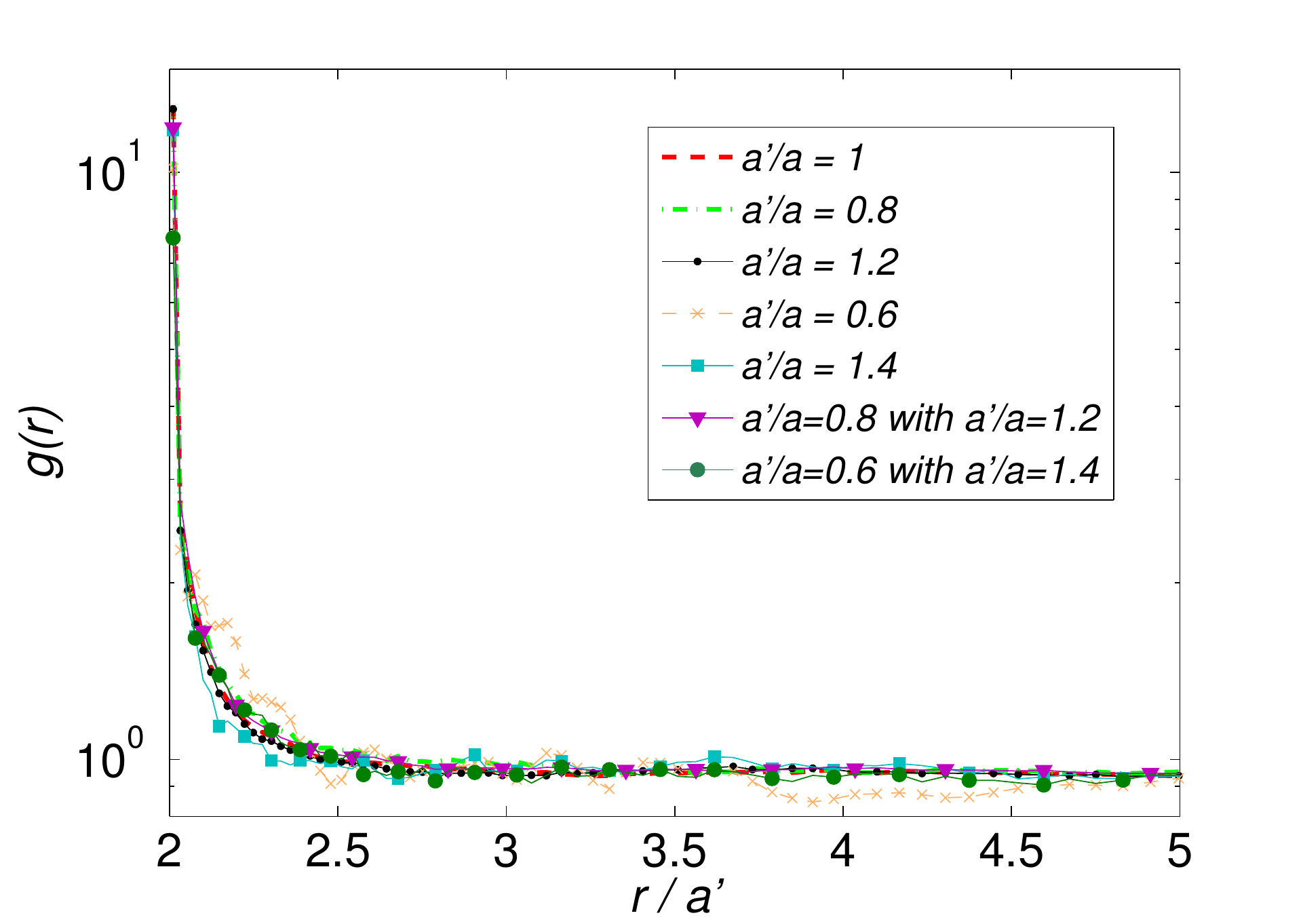}~~~~
\put(-185,110){{\large a)}}
{\includegraphics[width=.50\textwidth]{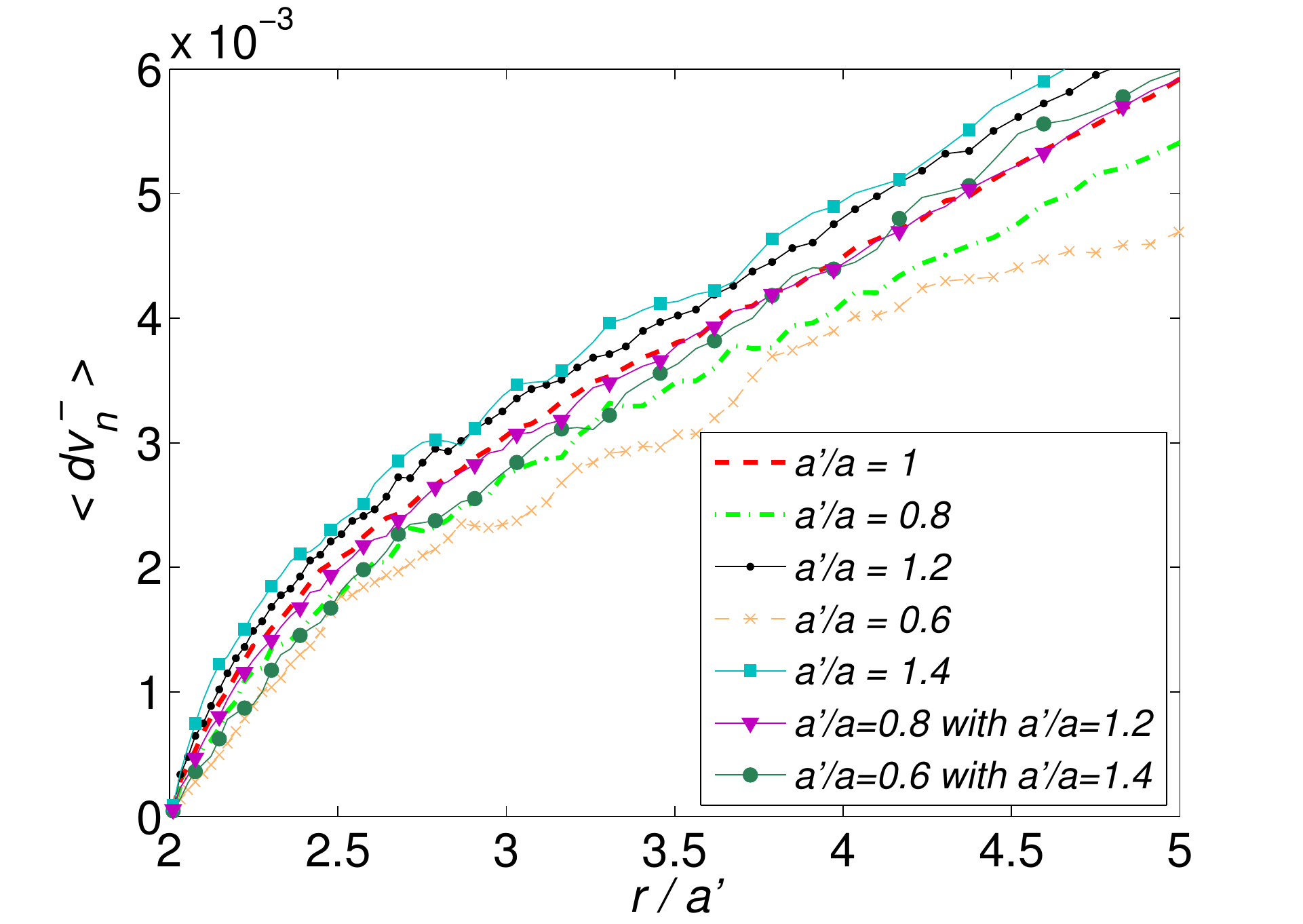}
\put(-180,110){{\large b)}}}
\includegraphics[width=.50\textwidth]{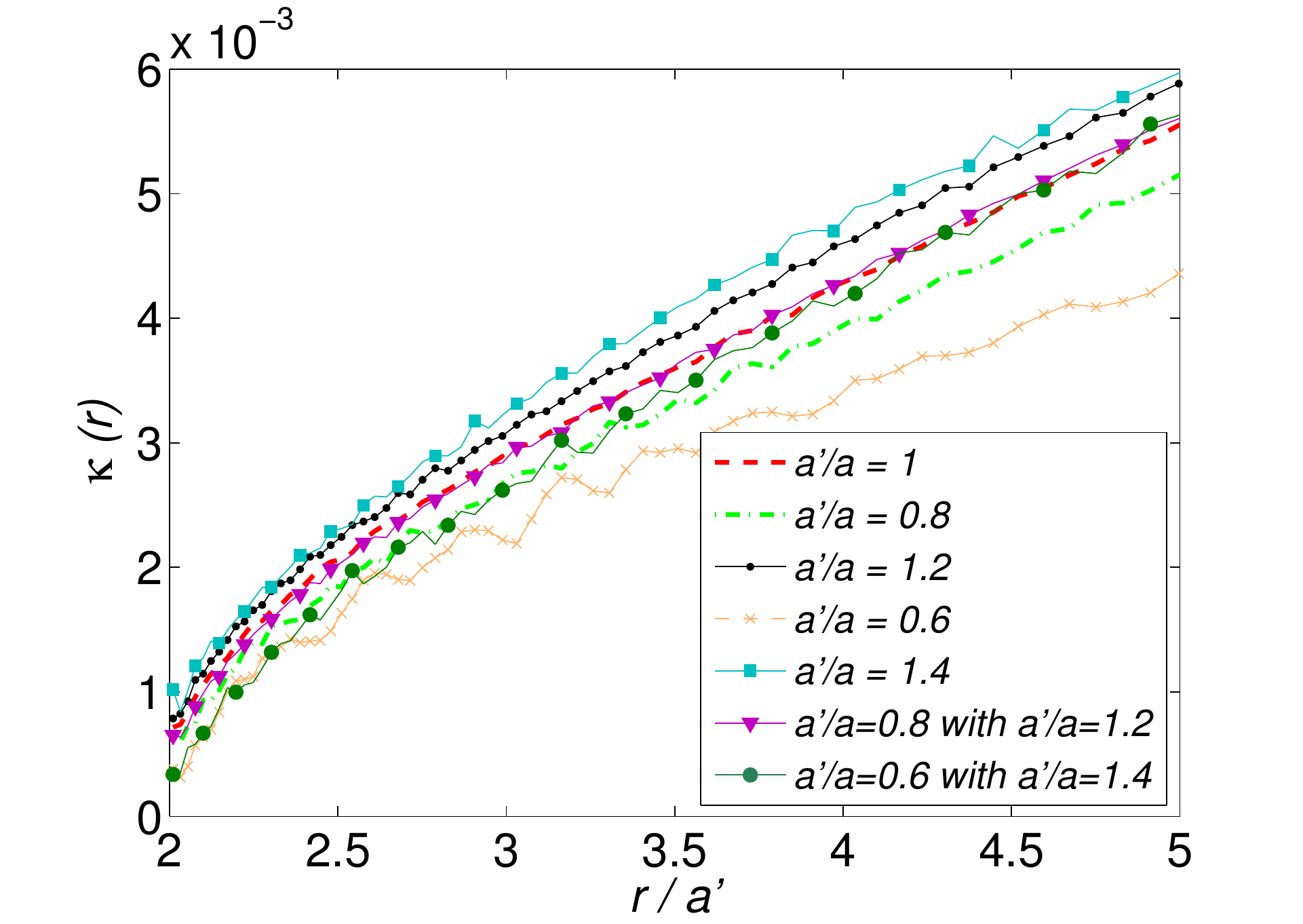}~~~~
\put(-185,110){{\large c)}}
{\includegraphics[width=.50\textwidth]{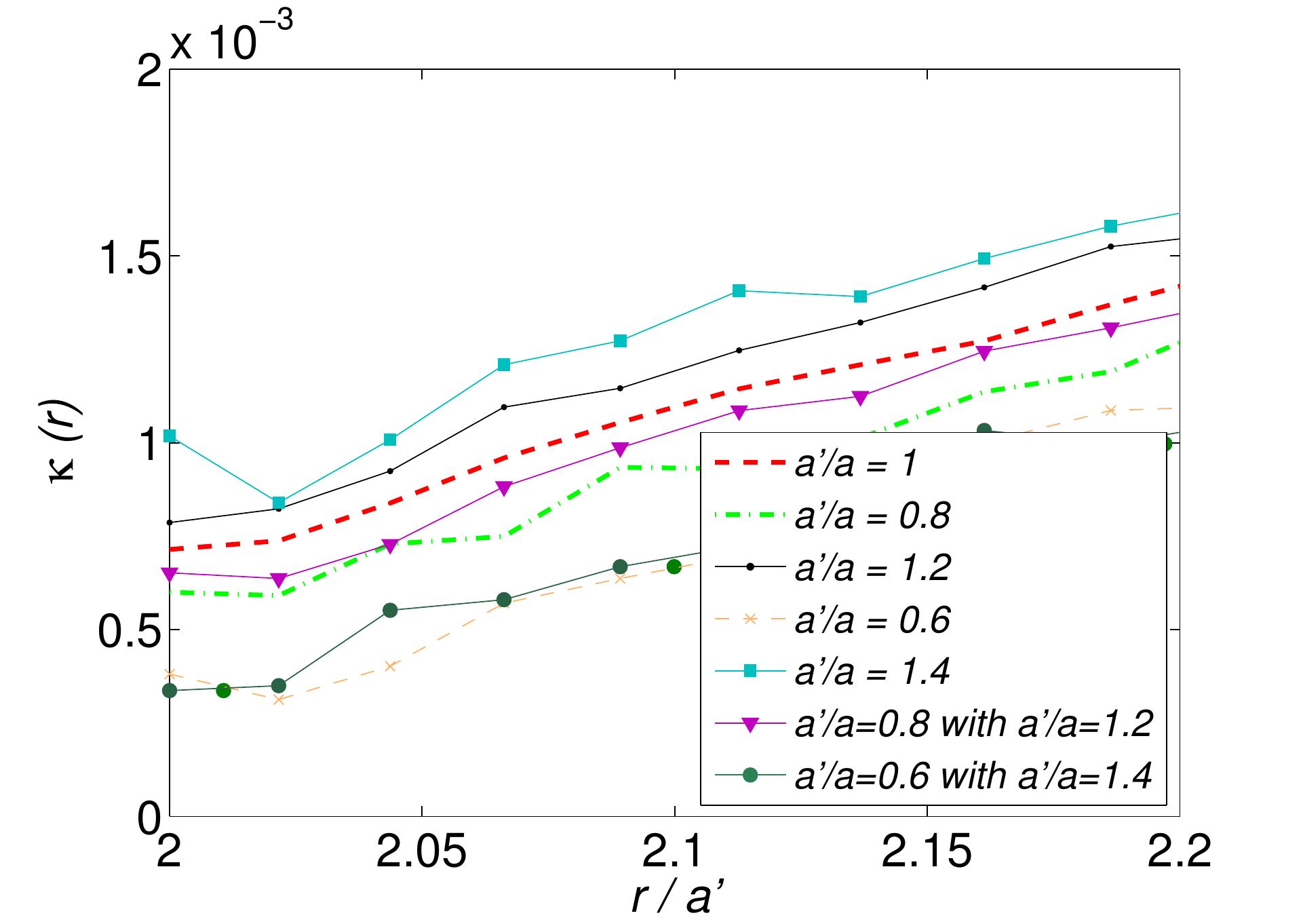}
\put(-180,110){{\large d)}}}
\caption{Radial distribution function $g(r)$ (a), average normal relative velocity $\langle dv_n^- \langle$ (b), collision kernel 
$\kappa(r)$ (c) and zoom of $\kappa(r)$ at contact (d), for $\sigma_a/(2a)=0.1$.}
\label{fig:coll}
\end{figure}

We then study particle-pair statistics. In particular we calculate the radial distribution function $g(r)$ and the averaged normal relative 
velocity between two approaching particles, $\langle dv_n^-(r)\rangle$, and finally the collision kernel $\kappa(r)$~\cite{sundaram1997}.\\
The radial distribution function $g(r)$ is an indicator of the radial separation among particle pairs. In a reference frame with origin at the 
centre of a particle, $g(r)$ is the average number of particle centers located in the shell of radius $r$ and thickness $\Delta r$, 
normalized with the number of particles of a random distribution. Formally the $g(r)$ is defined as 
\begin{equation}
\label{eqrdf}
g(r) = \frac{1}{4 \pi} \td{N_r}{r}\frac{1}{r^2 n_0},
\end{equation}
where $N_r$ is the number of particle pairs on a sphere of radius $r$, $n_0=N_p(N_p-1)/(2V)$ is the density of particle pairs in the volume 
$V$, with $N_p$ the number of particles. The value of $g(r)$ at distances of the order of the particle radius reveals the intensity of 
clustering; $g(r)$ tends to $1$ as $r \to \infty$, corresponding to a random (Poissonian) distribution. Here, we calculate it for 
pairs of particles with equal radii in the range $a'/a \in [0.6;1.4]$, and among particles of different sizes ($a'/a=0.8$ with $a'/a=1.2$ and 
$a'/a=0.6$ and $a'/a=1.4$). For each case, the radial distance $r$ is normalized by $a'$ or by the average between the radii of two 
approaching spheres. The radial distribution function is shown in figure~\ref{fig:coll}$(a)$. No appreciable differences between each curve 
can be observed. The $g(r)$ is found to drop quickly to the value of the uniform distribution (i.e. 1) at $r \sim 2.5a'$.\\
The normal relative velocity of a particle pair is instead obtained by projecting the relative velocity in the direction of the separation 
vector between the particles
\begin{equation}
dv_n(r_{ij}) = (\vec u_i - \vec u_j) \cdot \frac{(\vec r_i - \vec r_j)}{|(\vec r_i- \vec r_j)|} = (\vec u_i - \vec u_j) \cdot \frac{\vec r_{ij}}{|\vec r_{ij}|}
\end{equation}
(where $i$ and $j$ denote the two particles). This scalar quantity can be either positive (when two particles depart form each other) or 
negative (when they approach). Hence, the averaged normal relative velocity can be decomposed into 
$\langle dv_n(r) \rangle = \langle dv_n^+(r) \rangle + \langle dv_n^-(r) \rangle$. Here, we consider the absolute value of the mean negative 
normal relative velocity, shown in figure~\ref{fig:coll}$(b)$. We observe that larger particles approach with a slightly larger relative 
velocity $\langle dv_n^-(r)\rangle$ than smaller particles. This could be explained by looking at the probability density functions of the 
streamwise particle velocities shown in figure~\ref{fig:pdf}$(a)$. From this we see indeed that smaller particles can experience lower velocities 
with non-negligible probability, in comparison to larger particles.

Finally, figures~\ref{fig:coll}$(c),(d)$ report the collision kernel $\kappa(r)$ between particle-pairs. This is calculated as the 
product of the radial distribution function $g(r)$ and $\langle dv_n^-(r)\rangle$~\cite{sundaram1997}. At large seprataions, (i.e. 
$r/a' > 2.5$), we see that $\kappa(r)$ is fully dominated by the normal relative velocity. Around contact (i.e. $r/a' \simeq 2$) we see 
clearly that $\kappa(r)$ is higher for larger particles, see figure~\ref{fig:coll}$(d)$. The interesting result is found when looking at 
the collision kernels between particles of different sizes but equal concentration (within the suspension). For the case with $a'/a=0.8$ and 
$a'/a=1.2$, we see that $\kappa(r)$ is closer to that obtained for equal spheres with $a'/a=0.8$. Also for the case with $a'/a=0.6$ and 
$a'/a=1.4$, we see that $\kappa(r)$ is similar to that obtained for equal spheres with $a'/a=0.6$. This leads to the conclusion that collision 
statistics are dominated by the behavior of smaller particles.

%\begin{figure}
%\centering
%\includegraphics[width=.50\textwidth]{FIG/grr.eps}~~~~
%\put(-185,110){{\large a)}}
%{\includegraphics[width=.50\textwidth]{FIG/dvns.eps}
%\put(-180,110){{\large b)}}}
%\includegraphics[width=.50\textwidth]{FIG/krr.eps}~~~~
%\put(-185,110){{\large c)}}
%{\includegraphics[width=.50\textwidth]{FIG/zoomkr.eps}
%\put(-180,110){{\large d)}}}
%\caption{Radial distribution function $g(r)$ (a), average normal relative velocity $\langle dv_n^- \langle$ (b), collision kernel 
%$\kappa(r)$ (c) and zoom of $\kappa(r)$ at contact (d), for $\sigma_a/(2a)=0.1$.}
%\label{fig:coll}
%\end{figure}
%%%%%%%%%%%%%%%%%%%%%%%%%%%%%%%%%%%%%%%%%%%%%%%%%%%%%%%%%%%%%%%%%%%%%%%%%%%
\section{Final remarks}

We study numerically the behavior of monodisperse and polydisperse suspensions of rigid spheres in a turbulent channel flow. We consider 
suspensions with three different Gaussian distributions of particle radii (i.e. different standard deviations). The mean particle radius 
is equal to the reference radius of the monodisperse case. For the largest standard deviation, the ratio between largest and smallest 
particle radius is equal to $4$. We compare both fluid and particle statistics obtained for each case at a constant solid volume fraction 
$\phi=10\%$, hence, the total number of particles changes in each simulation.

The main finding of this work is that fluid and solid phase statistics for all polydisperse cases are similar to those obtained for the 
monodisperse suspension. This suggests that the key parameter in understanding the behavior of suspensions of rigid spheres in turbulent 
channel flows is the solid volume fraction $\phi$. Polydisperse suspensions with Gaussian distributions of particle sizes behave statistically 
in the same manner as a monodisperse suspension with equal volume fraction. This is probably not true for highly skewed size distributions.\\ 
Although results are similar, it is possible to observe small variations in the fluid and particle velocity fluctuations that are 
correlated to changes in the standard deviation $\sigma_a$ of the distribution. Concerning fluid velocity fluctuations we see that by increasing 
$\sigma_a$, these decrease at the centerline. The same is also found for particle velocity fluctuations. As $\sigma_a$ increases, 
larger particles are more likely found at the centerline and move almost unperturbed in the streamwise direction (hence also inducing smaller velocity 
fluctuations in the fluid phase). Particle velocity fluctuations are on the other hand found to increase with $\sigma_a$ close to the wall (in the 
viscous and buffer layers). This is probably related to the fact that for larger $\sigma_a$ smaller particles can penetrate 
more into this layer hence experiencing larger velocity fluctuations. Similar trends are observed for a smaller volume fraction of $2\%$.\\
Concerning the mean concentration of particles across the channel, we observe that the typical peak in proximity of the wall is 
smoothed for increasing $\sigma_a$. On the contrary, the mean concentration increases at the centerline at larger $\sigma_a$. 
Looking at the mean concentration profiles of each particle species (i.e. with different $a'$), we observe that all particles are uniformly 
distributed. For particles smaller than the reference ones, the near-wall peak moves closer to the wall, while for larger particles the peak 
is moved away from the wall and is smoothed for increasing $a'$.\\
We also calculated the Stokes number for each particles species. For the most extreme case, we found that there is an order of magnitude 
difference between $St_a'$ of larger and smaller particles. However, the mean Stokes number of the suspension is the same as that of the 
reference particles (i.e. the same as well as that of the monodisperse case). Hence, suspensions with same volume fraction 
and mean Stokes number behave statistically in a similar way. If a skewed distribution of particles was used, the mean Stokes number would 
change and probably different results to the monodisperse ones would be observed. This may have important implications for the modeling of 
particulate flows in channels.\\
Then, we looked at probability density functions, $pdf$s, of particle velocities as well as particles mean-squared displacements for each 
species for the most extreme case with $\sigma_a/(2a)=0.1$. Concerning the $pdf$s of streamwise velocity, we notice that smaller particles can 
penetrate more in the layers closer to the walls and hence also experience smaller velocities (wider left tail of the $pdf$). The opposite is 
of course found for particles with larger $a'$. The $pdf$s of wall-normal velocities are found to be extremely similar for all species, 
although the variance is just slightly increased for the smaller particles. This may be related to the fact that their Stokes number is smaller 
and hence they respond more quickly to velocity perturbations induced by turbulent eddies, hence reaching slightly larger velocities.\\
The mean-squared displacements of particles in the streamwise and spanwise directions, are similar for all species. Although 
larger particles are found to disperse less than smaller ones in the ballistic regime, the final diffusion coefficients are similar for all species. 
Finally, we studied particle-pair statistics by looking at the radial distribution function $g(r)$ and average normal relative velocity, 
$\langle dv_n^-(r)\rangle$, of approaching particles, as well as the resulting collision kernel, $\kappa(r)$. We found the interesting result 
that for pairs of particles with different sizes, the collision kernel $\kappa(r)$ is dominated by the behavior of smaller particles. 

We have therefore shown that in turbulent channel flows, polydisperse suspensions with gaussian distributions of sizes behave similarly to 
monodisperse suspensions, provided that: the volume fraction is constant; the mean Stokes number of the suspension is the same as that of the 
monodispersed particles. On the other hand, particles of different size lead to non trivial particle-pair statistics.

\section*{Acknowledgments}
%\begin{acknowledgments}
This work was supported by the European Research Council Grant No.\ ERC-2013-CoG-616186, TRITOS, from the Swedish Research Council (VR), through the 
Outstanding Young Researcher Award, and from the COST Action MP1305: \emph{Flowing matter}.
Computer time provided by SNIC (Swedish
National Infrastructure for Computing) and CINECA, Italy (ISCRA Grant FIShnET-HP10CQQF77).

\section*{References}

%\bibliography{biblio}

\end{document}